\documentclass[journal,12pt,draftclsnofoot,onecolumn]{IEEEtran}

\makeatletter
\def\ps@headings{%
\def\@oddhead{\mbox{}\scriptsize\rightmark \hfil \thepage}%
\def\@evenhead{\scriptsize\thepage \hfil \leftmark\mbox{}}%
\def\@oddfoot{}
\def\@evenfoot{}}
\def\blfootnote{\xdef\@thefnmark{}\@footnotetext}
\makeatother
\usepackage{graphicx} 
\usepackage{epsfig} 
\usepackage{color}
\usepackage{mathptmx} 
\usepackage{times} 
\usepackage{latexsym}
\usepackage{amsmath} 
\usepackage{amssymb}  
\usepackage{subfigure}
\usepackage{algorithmic}
\usepackage{setspace}
\usepackage[bottom]{footmisc}
\newtheorem{theorem}{Theorem}

\newtheorem{lemma}{Lemma}

\newtheorem{definition}{Definition}

\newtheorem{example}{Example}

\title{\huge \bf
Optimal Function Computation in Directed and Undirected Graphs
}             
                                                        
\author{\IEEEauthorblockN{Hemant Kowshik}\\
\IEEEauthorblockA{CSL and Department of ECE\\
University of Illinois Urbana-Champaign\\
Email: kowshik2@illinois.edu}\\ 
\and
\IEEEauthorblockN{P. R. Kumar}\\
\IEEEauthorblockA{CSL and Department of ECE\\
University of Illinois Urbana-Champaign\\
Email: prkumar@illinois.edu}
}
\begin{document}
\maketitle\blfootnote{This material is based upon work partially supported by AFOSR under Contract FA9550-09-0121, NSF under Contract No. CNS-1035378, Science \& Technology Center Grant CCF-0939370, Contract CNS-0905397, and Contract CNS-1035340, and USARO under Contract Nos. W911NF-08-1-0238 and W-911-NF-0710287. Any opinions, findings, and conclusions or recommendations expressed in this publication are those of the authors and do not necessarily reflect the views of the above agencies.}
\thispagestyle{empty}
\pagestyle{empty}
\begin{abstract}
We consider the problem of information aggregation in sensor networks, where one is interested in computing a \textit{function} of the sensor measurements. 
We allow for block processing and study in-network function computation in directed graphs and undirected graphs. We study how the structure of the function affects the encoding strategies, and the effect of interactive information exchange. Depending on the application, there could be a designated collector node, or every node might want to compute the function.

We begin by considering a directed graph $G = (\mathcal{V},\mathcal{E})$ on the sensor nodes, where the goal is to determine the optimal encoders on each edge which achieve function computation at the collector node. Our goal is to characterize the rate region in $\mathbf{R}^{|\mathcal{E}|}$, i.e., the set of points for which there exist feasible encoders with given rates which achieve zero-error computation for asymptotically large block length. We determine the solution for directed trees, specifying the optimal encoder and decoder for each edge. For general directed acyclic graphs, we provide an outer bound on the rate region by finding the disambiguation requirements for each cut, and describe examples where this outer bound is tight. 

Next, we address the scenario where nodes are connected in an undirected tree network, and every node wishes to compute a given symmetric Boolean function of the sensor data. Undirected edges permit interactive computation, and we therefore study the effect of interaction on the aggregation and communication strategies. We focus on sum-threshold functions, and determine the minimum worst-case total number of bits to be exchanged on each edge. The optimal strategy involves recursive in-network aggregation which is reminiscent of message passing. In the case of general graphs, we present a cut-set lower bound, and an achievable scheme based on aggregation along trees. For complete graphs, we prove that the complexity of this scheme is no more than twice that of the optimal scheme.
\end{abstract}
\doublespacing
\section{INTRODUCTION}
Sensor networks are composed of nodes with sensing, wireless communication and computation capabilities. These networks are designed for applications like fault monitoring, data harvesting and environmental monitoring; tasks which can be broadly classified as information aggregation. In these applications, one is interested only in computing some relevant \textit{function} of the sensor measurements. For example, one might want to compute the mean temperature for environmental monitoring, or the maximum temperature in fire alarm systems. This suggests moving away from a data-forwarding paradigm, and focusing on efficient in-network computation and communication strategies for the function of interest. This is particularly important since sensor nodes may be severely limited in terms of power and bandwidth, and can potentially generate enormous volumes of data.  

There are two possible architectures for sensor networks that one might consider. First, one could designate a single collector node/fusion center which seeks to compute the function. This goal is more appropriate for data harvesting and centralized fault monitoring. Alternately, one could suppose that every node in the network wants to compute the function. The latter goal can be viewed as providing situational awareness to each sensor node, which could be very useful in applications like distributed fault monitoring, adaptive sensing and sensor-actuator networks. For example, sensor nodes might want to modify their sampling rate depending on the value of the function. We will consider both these problems. 

In order to make progress on the general problem of computing functions of distributed data, we will study specific network topologies and some specific classes of functions. In this paper, we abstract out the medium access control problem associated with a wireless network, and view the network as a graph with edges representing noiseless links. The fundamental challenge is to exploit the structure of the particular function, so as to optimally combine transmissions at intermediate nodes. Thus, the problem of function computation could be regarded as being more general than finding the capacity of a wireless network. In our problem formulation, we consider the zero error block computation framework. We allow for nodes to accumulate a block of measurements and realize greater efficiency using block coding strategies.  However, we require the function to be computed with zero error for the block. To solve the problem under this framework, one needs to determine the optimal strategy for communication and computation, which includes determining the order in which nodes should transmit and the information that nodes must convey whenever they transmit. The strategy for computation may benefit from interactive information exchange between nodes, which presents an additional degree of freedom vis-a-vis the standard point-to-point communication set-up.

In Section \ref{sec_dir_graph}, we view the network as a directed graph with edges representing noiseless links. We thus consider the problem of general function computation in a directed graph $G = (\mathcal{V},\mathcal{E})$ with a designated collector. We focus specifically on strategies for combining information at intermediate nodes, and optimal codes for transmissions on each edge. We consider both the worst case and the average case complexity for zero error block computation with a joint probability distribution on the node measurements. Our goal is to characterize the rate region in $\mathbf{R}^{|\mathcal{E}|}$, i.e., the set of points for which there exist feasible encoders with given rates which achieve zero-error computation for large enough block length. In the case of tree graphs, we derive a necessary and sufficient condition for the encoder on each edge, which provides a complete characterization of the rate region. The extension of these results to directed acyclic graphs is more difficult. However, we provide an outer bound on the rate region by finding the disambiguation requirements for each cut, and describe examples where this outer bound is tight.

In Section \ref{sec_undir_graph}, we address the problem of computing symmetric Boolean functions in undirected graphs. The key difference from Section \ref{sec_dir_graph} is that we consider bidirectional links and study the benefit of interaction between nodes. We show how the approach described in Section \ref{sec_dir_graph}, together with ideas from communication complexity theory, can be synthesized to develop a theory of optimal computation of symmetric Boolean functions in undirected graphs. In the case of tree networks, each edge is a cut-edge, and this allows us to derive a lower bound on the number of bits exchanged on each edge, by considering an equivalent two node problem. Further, we show that a protocol of recursive in-network aggregation along with a smart interactive coding strategy, achieves this lower bound for the class of sum-threshold functions in tree networks. The optimal strategy has a simple structure that is reminiscent of message passing, where messages flow from the leaves towards an interior node, and then flow back from the interior node to the leaves. In the case of general graphs, we present a cut-set lower bound, and an achievable scheme based on aggregation along trees. For complete graphs, we show that the complexity of this scheme is no more than twice that of the optimal scheme. 
\section{RELATED WORK} 
In its simplest form, the problem of network function computation can be modeled as a problem of computation on graphs obtained by abstracting out the medium access control problem and channel noise. This problem is closely related to the network coding problem. Indeed, assuming independent measurements $x_i$ and  the identity function $f(x_1, x_2, \ldots, x_n) = (x_1, x_2, \ldots, x_n)$, we have the reverse of the multicast problem studied in \cite{AhlswedeYeung}. Computing a function of independent measurements is a network \textit{computation} problem as opposed to a network coding problem. In \cite{AppuFran}, the min-cut upper bound on the rate of computation is shown to be tight for the computation of divisible functions on tree graphs. In this paper, we generalize this result using a different approach. Further, the simplicity of the approach presented allows extensions to the case of general graphs and collocated networks.

The problem of worst-case block function computation was formulated in \cite{GiridharKumar}. The authors determine the maximum rate at which a symmetric function can be computed in a random network, given the constraints of the wireless medium. They identify two classes of symmetric functions namely \textit{type-sensitive} functions exemplified by Mean and Median, and \textit{type-threshold} functions, exemplified by Maximum and Minimum. The maximum rates for computation of type-sensitive and type-threshold functions in random planar networks are shown to be $\Theta(\frac{1}{\log n})$ and $\Theta(\frac{1}{\log \log n})$ respectively, for a network of $n$ nodes. A communication complexity approach was used to establish upper bounds on the rate of computation in collocated networks. Some extensions to the case of finite degree graphs are presented in \cite{GuptaShak}. 

In the study of the communication complexity of multi-party computation \cite{KushiNisan}, one seeks to minimize the number of bits that must be exchanged in the worst case between two nodes to achieve zero-error computation of a function of the node variables. The communication complexity of Boolean functions has been studied in \cite{Wegener}, \cite{OrlitskyElgamal}. Further, one can consider the \textit{direct-sum problem} \cite{KarchmerRazWigderson} where several instances of the problem are considered together to obtain savings. This block computation approach is used to compute the exact complexity of the Boolean AND function in \cite{AhlswedeCai}. In this paper, we considerably generalize this result, which allows us to derive optimal strategies for computing more general classes of symmetric Boolean functions in undirected tree networks. The optimal communication scheme is reminiscent of message passing algorithms which have been applied very effectively to the problems of computing marginals and probabilistic inference \cite{Kschischang}, \cite{AjiMceliece}.

An information-theoretic formulation of this problem combines the complexity of source coding of correlated sources with rate distortion, together with the complications introduced by the function structure; see \cite{GiridharKumar}. There is little or no work that addresses this most general framework. The problem of source coding with side information has been studied for the vanishing error case in \cite{WynerZiv}. This has been extended in \cite{OrlitskyRoche} to the case where the receiver desires to know a certain function $f(X,Y)$ of the single source $X$ and the side information $Y$; the authors determined the required capacity of the channel between the source and receiver to be the conditional graph entropy. However, the extension to larger networks has proved difficult. In Zero-error Information Theory, the problem of source coding with side information ensuring zero error for finite block length has been studied in \cite{Witsenhausen} and \cite{AlonOrlitsky}. The problem reduces to the task of coloring a probabilistic graph defined on the set of source samples. The minimum entropy of such a coloring approaches the graph entropy or Korner entropy, as the block length approaches infinity. Recently, the rate region for multi-round interactive function computation has been characterized for two nodes \cite{MaIshwar}, and for collocated networks \cite{MaGuptaIshwar}.

In this paper we do not address the problem of function computation in noisy networks. In \cite{Gallager}, the problem of computing parity in a collocated network in the presence of noise is considered. It is shown that $O(n \log \log n)$ bits suffice to achieve correct computation with high probability. This has been extended to random planar networks in  \cite{YingSrikantDullerud}, where the same $\log\log n$ factor of redundancy is shown to be sufficient. Remarkably, this factor was recently shown to be tight in \cite{Dutta}. 
\section{Function Computation in Directed Graphs}\label{sec_dir_graph}
In this section, we abstract out the medium access control problem associated with a wireless network, and view the network as a directed graph with edges representing essentially noiseless wired links between nodes. 
We formulate the problem of zero error function computation on graphs. We suppose that there is a joint probability distribution on the node measurements, and allow nodes to realize greater efficiency by using block codes. We will consider both the worst case and the average case complexity for zero error block computation. Given a graph, the problem we address is to determine the set of rates on the edges which will allow zero error function computation for a large enough block length. In essence, we are exploring the interaction between the function structure and the structure of the graph; how information needs to be routed and combined at intermediate nodes to achieve certain rate vectors. 

In Section \ref{sec_two_node_dir}, we begin with the two node problem. We compute the number of bits that node $v_X$ needs to communicate to node $v_Y$ so that the latter can compute a function $f(X,Y)$ with zero error. For correct function computation, an encoder must disambiguate certain pairs of source symbols of node $v_X$, on which the function disagrees. We show by explicit construction of a code that this necessary condition is in fact sufficient. This yields the optimal alphabet and we calculate the minimum worst case and average case complexity, with the latter obtained by Huffman coding over the optimal alphabet. In Section \ref{sec_tree}, we extend this result to directed trees with the collector as root, exploiting the fact that each edge is a cut-edge. This yields the optimal alphabet for each edge, and we separately optimize the encoders for the worst case and the average case. Thus the rate region consists of all rate points dominating a single point that is coordinate-wise optimal. 

In Section \ref{sec_DAG}, we consider directed acyclic graphs. A key difference from the tree case is the presence of multiple paths to route the data, which present different opportunities to combine information at intermediate nodes. We arrive at an outer bound to the rate region by finding the disambiguation requirements for each cut of the directed graph. This outer bound is not always tight, as we show in Example \ref{ex_arithmetic_sum}. However, for the worst case computation of finite field parity, and the  maximum or minimum functions, the outer bound is shown to be indeed tight. Further, the only extreme points of the rate region are rate points corresponding to activating only a tree subset of edges. 
\subsection{Two Node Setting}\label{sec_two_node_dir}
\subsubsection{Worst case complexity}
We begin by considering the simple two node problem. Suppose nodes $v_X$ and $v_Y$ have measurements $x \in \mathcal{X}$ and $y \in \mathcal{Y}$, where the alphabets $\mathcal{X}$ and $\mathcal{Y}$ are finite sets. 
Node $v_X$ needs to optimally communicate its information to node $v_Y$ so that a function $f(x,y)$, which takes values in $\mathcal{D}$, can be computed at $v_Y$ with zero error. We do not consider the case where $v_X$ and $v_Y$ interactively compute the function. Thus node $v_X$ has an encoder $\mathcal{C}:\mathcal{X} \rightarrow \{0,1\}^{*}$, which maps its measurement $x$ to the codeword $\mathcal{C}(x)$, and node $v_Y$ has a decoder $g:\{0,1\}^{*} \times \mathcal{Y} \rightarrow \mathcal{D}$ which maps the received codeword $\mathcal{C}(x)$ and its own measurement $y$ to a function estimate, $g(\mathcal{C}(x), y)$. The set of all possible codewords is called the codebook, denoted by $\mathcal{C}(\mathcal{X})$
\begin{definition}[Feasible Encoder]
An encoder $\mathcal{C}$ is \textit{feasible} if there exists a decoder $g:\{0,1\}^{*} \times \mathcal{Y} \rightarrow \mathcal{D}$ such that $g(\mathcal{C}(x), y) = f(x,y)$  for all $(x,y) \in \mathcal{X} \times \mathcal{Y}$. Thus, a feasible encoder is one that achieves error-free function computation.
\end{definition}
\begin{theorem}[Characterization of Feasible Encoders]\label{nec_condn_1}
An encoder $\mathcal{C}$ is feasible if and only if given $x^1, x^2 \in \mathcal{X}$, $\mathcal{C}(x^1) = \mathcal{C}(x^2)$ implies $f(x^1, y) = f(x^2, y)$ for all $y \in \mathcal{Y}$.
\end{theorem}
\textbf{Proof:} By definition, if $\mathcal{C}$ is a feasible encoder, then there exists a corresponding decoder $g$ such that $g(\mathcal{C}(x^1),y) = f(x^1, y)$ and $g(\mathcal{C}(x^2),y) = f(x^2, y)$, for all $y \in \mathcal{Y}$. Further, if $\mathcal{C}(x^1) = \mathcal{C}(x^2)$, we have $f(x^1, y) = f(x^2, y)$ for all $y \in \mathcal{Y}$.

To prove the converse, we need to construct a decoding function $g:\{0,1\}^{*} \times \mathcal{Y} \rightarrow \mathcal{D}$. For each codeword $C^{*}$ in the codebook, define $\mathcal{C}^{-1}(C^{*}) := \{x \in \mathcal{X}: \mathcal{C}(x) = C^{*}\}$. For fixed $y \in \mathcal{Y}$ and fixed codeword $C^{*} \in \mathcal{C}(\mathcal{X})$, the decoder mapping is given by $g(C^{*},y) := f(x^{nom}(C^{*}),y)$ for any arbitrary $x^{nom}(C^{*}) \in \mathcal{C}^{-1}(C^{*})$. We show that this decoder works for any fixed $x$ and $y$ . Indeed, $ g(\mathcal{C}(x),y) = f(x^{nom},y)$ where $x^{nom} \in \mathcal{C}^{-1}(\mathcal{C}(x))$. Thus, $\mathcal{C}(x^{nom}) = \mathcal{C}(x)$ and by assumption $f(x^{nom},y) = f(x,y)$. Hence, $g(\mathcal{C}(x),y) = f(x^{nom},y) = f(x,y)$ for all $y \in \mathcal{Y}$. $\Box$

Any feasible encoder $\mathcal{C}$ can be viewed as partitioning the set $\mathcal{X}$ into $\Pi(\mathcal{C}) := \{S_1, S_2, \ldots, S_k\}$ such that for $x^{1} \in C_i, x^2 \in C_j$, we have $\mathcal{C}(x^1) = \mathcal{C}(x^2)$ if and only if $i = j$. Define an equivalence relation ``$\leftrightarrow$'' between $x^1, x^2 \in \mathcal{X}$ by:
\begin{displaymath}
x^1 \leftrightarrow x^2 \textrm{ if and only if } f(x^1, y) = f(x^2, y) \textrm{ for all } y \in \mathcal{Y}.
\end{displaymath}
Consider the encoder $\mathcal{C}^{OPT}$ which assigns a distinct codeword to each resulting equivalence class. Clearly, $\mathcal{C}^{OPT}$ is a feasible encoder, since $\mathcal{C}^{OPT}(x^1) = \mathcal{C}^{OPT}(x^2)$ implies $x^1 \leftrightarrow x^2$, and hence $f(x^1, y) = f(x^2, y)$ for all $y \in \mathcal{Y}$. $\mathcal{C}^{OPT}$ is optimal in the sense that any other feasible encoder $\mathcal{C}$ must have at least as many codewords as $\mathcal{C}^{OPT}$:
\begin{theorem}[Optimality of $\mathcal{C}^{OPT}$]\label{opt_encoder}
Let $\Pi(\mathcal{C}^{OPT}) := \{S^{OPT}_1, S_2^{OPT}, \ldots, S_k^{OPT}\}$ be the partition of $\mathcal{X}$ generated by $\mathcal{C}^{OPT}$, and let $\Pi(\mathcal{C}) := \{S_1, S_2, \ldots, S_l\}$ be the partition of $\mathcal{X}$ generated by any other feasible encoder $\mathcal{C}$. Then, \\
(i) $\Pi(\mathcal{C})$ must be a finer partition than $\Pi(\mathcal{C}^{OPT})$.\\
(ii) The minimum number of bits that node $v_X$ needs to communicate is $\lceil \log |\Pi(\mathcal{C}^{OPT})| \rceil$. 
\end{theorem}
\textbf{Proof: } First we claim that any subset $S_i \in \Pi(\mathcal{C})$ can have nonempty intersection with exactly one subset $S^{OPT}_j \in \Pi(\mathcal{C}^{OPT})$. Suppose not. Then there exist $x^1, x^2 \in S_{i}$ such that $x^1 \in S^{OPT}_{j_1}$ and $x^2 \in S^{OPT}_{j_2}$. Since $\mathcal{C}(x^1) = \mathcal{C}(x^2)$, by Theorem \ref{nec_condn_1}, we must have $f(x^1,y) = f(x^2, y)$ for all $y \in \mathcal{Y}$. However, by construction of $\mathcal{C}^{OPT}$, $x^1$ and $x^2$ must belong to distinct equivalence classes i.e., $x^1 \nleftrightarrow x^2$. Hence, there exists $y^*$ such that $f(x^1, y^*) \neq f(x^2, y^*)$, which is a contradiction. This shows that the partition generated by any encoder $\mathcal{C}$ must be a further subdivision of the partition generated by $\mathcal{C}^{OPT}$, i.e., finer than $\Pi(\mathcal{C}^{OPT})$. So node $v_X$ needs to communicate at least $\lceil \log |\Pi(\mathcal{C}^{OPT})| \rceil$ bits. $\Box$

We can extend this to the case where $v_X$ collects a block of $N$ measurements $\underline{x} = (x_1, x_2, \ldots, x_N) \in \mathcal{X}^{N}$, and $v_Y$ collects a block of $N$ measurements $\underline{y} = (y_1, y_2, \ldots, y_N) \in \mathcal{Y}^{N}$. We want to find a block encoder $\mathcal{C}^{N}:\mathcal{X}^N \rightarrow \{0,1\}^{*}$ so that the vector function $f^{(N)}(\underline{x}, \underline{y}) = (f(x_1, y_1), \ldots, f(x_N, y_N))$ can be computed without error, 
for all $\underline{x} \in \mathcal{X}^{N}, \underline{y} \in \mathcal{Y}^{N}$. All the above results carry over to the error-free block computation case. As before, we define an equivalence $\leftrightarrow$ between $\underline{x}^1, \underline{x}^2 \in \mathcal{X}^N$ if $f^{(N)}(\underline{x}^1, \underline{y}) = f^{(N)}(\underline{x}^2, \underline{y})$ for all $\underline{y} \in \mathcal{Y}^N$. The optimal encoder $\mathcal{C}^{N,OPT}$ is once again obtained by assigning distinct codewords to each equivalence class. Since we are stringing together $N$ independent instances, we have $|\Pi(\mathcal{C}^{N,OPT})| = |\Pi(\mathcal{C}^{OPT})|^{N}$. Hence the minimum number of bits per computation that node $v_X$ needs to communicate is $\frac{\lceil N \log|\Pi(\mathcal{C}^{OPT})| \rceil}{N}$ which converges to $\log |\Pi(\mathcal{C}^{OPT})|$ as $N \rightarrow \infty$.
\subsubsection{Average case complexity}
Suppose now that the measurements $X$, $Y$ are drawn from the joint probability distribution $p(X,Y)$, with the goal being to minimize the average number of bits that need to be communicated, i.e., the \textit{average case complexity}. 
\begin{definition}[Feasible Encoder]
An encoder $\mathcal{C}:\mathcal{X} \rightarrow \{0,1\}^{*}$ is \textit{feasible} if there exists a decoder  $g:\{0,1\}^{*} \times \mathcal{Y} \rightarrow \mathcal{D}$ such that $g(\mathcal{C}(x), y) = f(x,y)$ for all $\{(x,y) \in \mathcal{X} \times \mathcal{Y}: p(x,y) >0\}$.
\end{definition}
\begin{theorem}\label{nec_condn_2}
An encoder $\mathcal{C}$ is feasible if and only if, given $x^1, x^2 \in \mathcal{X}$,  $\mathcal{C}(x^1) = \mathcal{C}(x_2)$ implies $f(x_1, y) = f(x_2, y)$ for $\{y \in \mathcal{Y}: p(x_1,y)p(x_2,y)>0\}$.
\end{theorem}
\textbf{Proof: }By definition, if $\mathcal{C}$ is a feasible encoder, then there exists a corresponding decoder $g$ such that $g(\mathcal{C}(x^1),y) = f(x^1, y)$ and $g(\mathcal{C}(x^2),y) = f(x^2, y)$, for all $\{y \in \mathcal{Y}: p(x^1,y)p(x^2,y)>0\}$. Further, if $\mathcal{C}(x^1) = \mathcal{C}(x^2)$, we have $f(x^1, y) = f(x^2, y)$ for  $\{y \in \mathcal{Y}: p(x^1,y)p(x^2,y)>0\}$.

To prove the converse, we need to construct a decoding function $g:\{0,1\}^{*} \times \mathcal{Y} \rightarrow \mathcal{D}$. For each codeword $\mathcal{C}^{*}$ in the codebook, define $\mathcal{C}^{-1}(C^{*}) := \{x \in \mathcal{X}: \mathcal{C}(x) = C^{*}\}$. For fixed $y \in \mathcal{Y}$ and fixed codeword $\mathcal{C}^{*} \in \mathcal{C}(\mathcal{X})$, the decoder mapping is given by $g(C^{*},y) := f(x^{nom}(C^{*},y),y)$ for any arbitrary $x^{nom}(C^{*},y) \in \mathcal{C}^{-1}(C^{*})$ with $p(x^{nom}(C^{*},y), y) > 0$. We show that this decoder works for any fixed $x$ and $y$ with $p(x,y)>0$. Indeed, $ g(\mathcal{C}(x),y) = f(x^{nom},y)$ where $x^{nom} \in \mathcal{C}^{-1}(\mathcal{C}(x))$ with $p(x^{nom},y) > 0$. Thus, $\mathcal{C}(x^{nom}) = \mathcal{C}(x)$ and by assumption $f(x^{nom},y) = f(x,y)$ since $p(x^{nom},y)p(x,y)>0$. Hence, $g(\mathcal{C}(x),y) = f(x^{nom},y) = f(x,y)$. $\Box$

We now define ``$x^1 \leftrightarrow x^2$'' when $f(x^1, y) = f(x^2, y)$ for $\{y \in \mathcal{Y}: p(x^1,y)p(x^2,y)>0\}$. Now the $\leftrightarrow$ relation is reflexive and symmetric, but not necessarily transitive. 
However, if $p(x,y) > 0$ for all $(x,y) \in \mathcal{X} \times \mathcal{Y}$, then $\leftrightarrow$ is an equivalence relation. We can construct an encoder $\mathcal{C}^{OPT}$ which assigns a distinct codeword to each equivalence class. Let $\Pi(\mathcal{C}^{OPT}) := \{S^{OPT}_1, S_2^{OPT}, \ldots, S_k^{OPT}\}$ be the partition of $\mathcal{X}$ generated by $\mathcal{C}^{OPT}$. Analogous to Theorem \ref{opt_encoder}, we can show that the encoder $\mathcal{C}^{OPT}$ has the \textit{optimal alphabet} $\mathcal{A}$, with the probability distribution vector $\underline{q} = \{q_1, q_2, \ldots, q_k\}$ where
$q_i := \sum_{x \in S^{OPT}_{i}}\sum_{y \in \mathcal{Y}}p(x,y)$.

Once the optimal alphabet is fixed, the optimal code $\mathcal{C}^{OPT}$ is the \textit{binary Huffman code} for the probability vector $\underline{q}$. Since the Huffman code has an average code length within one bit of the entropy, 
\begin{displaymath}
H(q_1, q_2, \ldots, q_k) \leq E[l(\mathcal{C}^{OPT})] \leq H(q_1, q_2, \ldots, q_k) + 1.
\end{displaymath}

The extension to the case where nodes $v_X$,$v_Y$ collect a block of $N$ i.i.d. measurements is straightforward. 
The optimal alphabet is $\mathcal{A}^{N}$, which has the product distribution $\underline{q}^{N}$.  The optimal encoder is obtained via the Huffman code for the optimal alphabet. Its expected length satisfies
\begin{displaymath}
\frac{H(\underline{q}^{N})}{N} \leq \frac{E[l(\mathcal{C}^{N,OPT})]}{N} \leq \frac{H(\underline{q}^{N}) + 1}{N}.
\end{displaymath}
Hence the minimum number of bits per computation that node $v_X$ needs to communicate converges to $H(\underline{q})$ as $N \rightarrow \infty$.
\subsection{Function Computation in Directed Trees}\label{sec_tree}
Let us now consider computation on a \textit{tree graph}. Consider a directed tree $G = (\mathcal{V}, \mathcal{E})$ with nodes $\mathcal{V} := \{v_1, v_2, \ldots, v_n\}$ and root node $v_1$. Edges represent communication links, so that node $v_j$ can transmit to node $v_i$ if $(v_j, v_i) \in \mathcal{E}$. Each node $v_i$ makes a measurement $x_{i} \in \mathcal{X}_i$, and the collector node $v_1$ wants to compute a function $f(x_1, x_2, \ldots, x_n)$ with no error. We seek to minimize the worst case complexity on each edge. 

For each node $i$, let $\pi(v_i)$ be the unique node to which node $i$ has an outgoing edge, and let $\mathcal{N}^{-}(v_i) := \{v_j \in \mathcal{V}: (v_j, v_i) \in \mathcal{E}\}$. The \textit{height} of a node $v_i$ is the length of the longest directed path from a leaf node to $v_i$. Define the \textit{descendant set} $D(v_i)$ to be the subset of nodes in $\mathcal{V}$ from which there exist directed paths to node $v_i$. The graph induced on $D(v_i)$ is a tree with node $v_i$ as root. Each node transmits exactly once and the computation proceeds in a bottom-up fashion, starting from the leaf nodes and proceeding up the tree. 

Each leaf node $v_i$ has an encoder $\mathcal{C}_i: \mathcal{X}_i \rightarrow \{0,1\}^{*}$ that maps its measurement $x_i$ to a codeword $\mathcal{C}_{i}(x_i)$ which is transmitted on the edge $(v_i, \pi(v_i))$. Each non-leaf node $v_j$ for $j\neq 1$ has an encoder $\mathcal{C}_{j}$ which maps its measurement $x_j$ as well as the codewords received from $\mathcal{N}^{-}(v_j)$, to a codeword transmitted on the edge $(v_j, \pi(v_j))$. Thus the computation proceeds in a bottom-up fashion. Let $C_i$ denote the codeword transmitted by node $v_i$, and $C_{S} := \{C_i: v_i \in S\}$ denote the set of codewords transmitted by nodes in $S$. 
\begin{definition}
A set of encoders $\{\mathcal{C}_i: 2 \leq i \leq n\}$ is said to be feasible if there is a decoding function $g_1$ at the collector node $v_1$ such that $g(x_1, \mathcal{C}_{\mathcal{N}^{-}(v_1)}) = f(x_1, x_2, \ldots, x_n)$ for all $(x_1, x_2, \ldots, x_n) \in \mathcal{X}_1 \times \mathcal{X}_2 \times \ldots \times \mathcal{X}_n$.
\end{definition}
\begin{lemma}\label{lem_nec}
If a set of encoders $\{\mathcal{C}_i: 2 \leq i \leq n\}$ is feasible, then the encoder $\mathcal{C}_i$ at node $v_i$ must separate\footnote{Node $v_i$ does not have access to $x_{D(v_i)}$ directly but only the codewords received from $\mathcal{N}^{-}(v_i)$. When we say that the encoder $\mathcal{C}_{i}$ must separate $x_{D(v_i)},\tilde{x}_{D(v_i)}$, we are considering $\mathcal{C}_i$ as an implicit function of $x_{D(v_i)}$.} $x^{1}_{D(v_i)} \in \mathcal{X}_{D(v_i)}$ from $x^{2}_{D(v_i)} \in \mathcal{X}_{D(v_i)}$, if there exists an assignment $x^{*}_{\mathcal{V} \setminus D(v_i)}$ such that $f(x^{1}_{D(v_i)}, x^{*}_{\mathcal{V} \setminus D(v_i)}) \neq f(x^{2}_{D(v_i)}, x^{*}_{\mathcal{V}\setminus D(v_i)})$.
\end{lemma}
\textbf{Proof: }The removal of edge $(v_i, \pi(v_i))$ separates the graph into two disconnected subtrees $D(v_i)$ and $\mathcal{V}\setminus D(v_i)$. We combine all the nodes in $D(v_i)$ into a supernode $v_{\alpha}$, and all the nodes in $\mathcal{V}\setminus D(v_i)$ into a supernode $v_{\beta}$. The result now follows from Theorem \ref{nec_condn_1}. $\Box$

To prove the converse, we explicitly define the encoders $\mathcal{C}_2, \mathcal{C}_3, \ldots, \mathcal{C}_n$ and a decoding function $g$, and prove that it achieves correct function computation. 
Define the alphabet for encoder $\mathcal{C}_i$ on edge $(v_i, \pi(v_i))$ as,
\begin{multline*}
\mathcal{A}_{i} := \{h_i:\mathcal{X}_{\mathcal{V}\setminus D(v_i)} \rightarrow \mathcal{D} \textrm{ s. t. } \exists x^{*}_{D(v_i)} \in \mathcal{X}_{D(v_i)}, \\
h_i(x_{\mathcal{V}\setminus D(v_i)}) = f(x^{*}_{D(v_i)}, x_{\mathcal{V}\setminus D(v_i)}) \quad \forall x_{\mathcal{V}\setminus D(v_i)} \in \mathcal{X}_{\mathcal{V}\setminus D(v_i)}\}.
\end{multline*}
Thus codewords sent by node $v_i$ can be viewed as \textit{normal forms} on variables $X_{\mathcal{V} \setminus D(v_i)}$, or as partial functions on $X_{\mathcal{V} \setminus D(v_i)}$. \\
\textbf{Encoder at node $v_i$:} On receiving the codeword corresponding to $h_j: \mathcal{X}_{\mathcal{V}\setminus D(v_j)} \rightarrow \mathcal{D}$, on incoming edge $(v_j, v_i)$, node $v_i$ assigns nominal values, $x^{nom}_{D(v_j)}$ to variables $X_{D(v_j)}$ such that
\begin{equation}\label{equal_one}
f(x^{nom}_{D(v_j)}, x_{\mathcal{V} \setminus D(v_j)}) = h_j(x_{\mathcal{V} \setminus D(v_j)})  \quad \forall x_{\mathcal{V} \setminus D(v_j)} \in \mathcal{X}_{\mathcal{V} \setminus D(v_j)}.
\end{equation}
Given nominal values for all nodes in $D(v_i) \setminus \{v_i\}$, and its own measurement $x_i$, node $v_i$ substitutes these values to obtain a function $h_i: \mathcal{X}_{\mathcal{V} \setminus D(v_i)} \rightarrow \mathcal{D}$ such that 
\begin{displaymath}
h_i(x_{\mathcal{V} \setminus D(v_i)}) = f(x^{nom}_{D(v_i) \setminus \{v_i\}}, x_i, x_{\mathcal{V} \setminus D(v_i)}) \textrm{ for all } x_{\mathcal{V} \setminus D(v_i)} \in \mathcal{X}_{\mathcal{V}
 \setminus D(v_i)}. 
\end{displaymath}

If $v_i \neq v_1$, node $v_i$ then transmits the codeword $\mathcal{C}_i$ corresponding to  function $h_i \in \mathcal{A}_i$ on the edge $(v_i, \pi(v_i))$ . \\
\textbf{Decoding function $g$:}
The collector node $v_1$ assigns nominal values to the variables $X_{D(v_1) \setminus \{v_1\}}$. The decoding function $g$ is given by
$g(x_1, C_{\mathcal{N}^{-}(v_1)}) := h_1 = f(x_1, x^{nom}_{D(v_1) \setminus \{v_1\}})$. 
\begin{theorem}
Let $x_1^{fix}, x_2^{fix}, \ldots, x_n^{fix}$ be any fixed assignment of node values. Let the encoders at node $v_2, v_3, \ldots, v_n$ be as above. Then function $h_i$ computed by node $v_i$ is,
\begin{displaymath}
h_i(x_{\mathcal{V}\setminus D(v_i)}) = f(x^{fix}_{D(v(i))}, x_{\mathcal{V}\setminus D(v_i)}) \qquad \forall x_{\mathcal{V}\setminus D(v_i)} \in \mathcal{X}_{\mathcal{V}\setminus D(v_i)}. 
\end{displaymath}
Consequently the decoding function $g$ satisfies $g(x^{fix}_1, \mathcal{C}_{\mathcal{N}^{-}(v_1)}) = f(x^{fix}_1, x^{fix}_2, \ldots, x^{fix}_n)$. 
\end{theorem}
\textbf{Proof: }The proof proceeds by induction. The theorem is trivially true for all leaf nodes $v_i$, since by assumption $h_i(x_{\mathcal{V} \setminus D(v_i)}) = f(x^{fix}_{v_i}, x_{\mathcal{V} \setminus D(v_i)})$ for all $ x_{\mathcal{V} \setminus D(v_i)} \in \mathcal{X}_{\mathcal{V} \setminus D(v_i)}$. Suppose it is true for all nodes with height less than $\kappa$. Consider a node $v_i$ with height $\kappa$. All the nodes in $\mathcal{N}^{-}(v_i)$ must have height less than $\kappa$. On receiving the codeword corresponding to $h_j$ on edge $(v_{j}, v_i)$, node $v_i$ assigns nominal values to variables in $X_{D(v_{j})}$ so that (\ref{equal_one}) is satisfied. From the induction assumption, we have
\begin{equation}\label{equal_two}
h_{j}(x_{\mathcal{V}\setminus D(v_{j})}) = f(x^{fix}_{D(v_j)}, x_{\mathcal{V}\setminus D(v_j)})  
\quad \forall x_{\mathcal{V}\setminus D(v_j)} \in \mathcal{X}_{\mathcal{V}\setminus D(v_j)}.
\end{equation}
Since (\ref{equal_two}) is true for all $v_j \in \mathcal{N}^{-}(v_i)$, we can simultaneously substitute the nominal values $x^{nom}_{D(v_i) \setminus \{v_i\}}$ for the variables $X_{D(v_i) \setminus \{v_i\}}$ and the value $x^{fix}_{i}$ for the variable $X_{\{v_i\}}$, to obtain a function $h_i$ satisfying
\begin{eqnarray}
h_i(x_{\mathcal{V}\setminus D(v_i)}) & \hspace{-0.1in}= & f(x^{nom}_{D(v(i)) \setminus \{v_i\}}, x^{fix}_{v_i}, x_{\mathcal{V}\setminus D(v_i)}) \quad \forall x_{\mathcal{V}\setminus D(v_i)} \nonumber \\
& \hspace{-0.1in}= & f(x^{fix}_{D(v(i))}, x_{\mathcal{V}\setminus D(v_i)}) \qquad \qquad \forall x_{\mathcal{V}\setminus D(v_i)}, \label{equal_three}
\end{eqnarray}
where (\ref{equal_three}) follows from (\ref{equal_one}) and (\ref{equal_two}). This establishes the induction step and completes the proof. For the special case of the collector node $v_i$, we have
\begin{displaymath}
g(x^{fix}_1, \mathcal{C}_{\mathcal{N}^{-}(v_1)}) = h_1 = f(x^{fix}_{D(v_1)}) = f(x^{fix}_1, x^{fix}_2, \ldots, x^{fix}_n).
\end{displaymath}
Since this is true for every fixed assignment of the node values, we can achieve error-free computation of the function. Hence the set of encoders described above is feasible. $\Box$

For node $v_i$, consider the equivalence relation ``$\leftrightarrow_{i}$'' where $x^{1}_{D(v_i)} \leftrightarrow_{i} x^{2}_{D(v_i)}$ if $f(x^{1}_{D(v_i)}, x_{\mathcal{V} \setminus D(v_i)}) = f(x^{2}_{D(v_i)}, x_{\mathcal{V} \setminus D(v_i)})$ for all $x_{\mathcal{V} \setminus D(v_i)} \in \mathcal{X}_{\mathcal{V} \setminus D(v_i)}$. It is easy to check that the equivalence classes generated by $\leftrightarrow_{i}$ are captured exactly by the alphabet $\mathcal{A}_i$. Thus the above encoders use exactly the optimal alphabet. Hence, the minimum worst case complexity for encoder $\mathcal{C}_i$ is $\lceil \log(|\mathcal{A}_i|) \rceil$ on the edge $(v_i, \pi(v_i))$.

The extension to the case where node $v_i$ collects a block of $N$ independent measurements $\underline{X}_i \in \mathcal{X}_i^{N}$, and the collector node $v_1$ wants to compute the vector function  $f^{(N)}(\underline{X}_1, \underline{X}_2, \ldots, \underline{X}_n)$, is straightforward. We can thus achieve a minimum worst case complexity arbitrarily close to $\log|\mathcal{A}_i|$ bits for encoder $\mathcal{C}_i$. It should be noted that 
the minimum worst case complexity of encoder $\mathcal{C}_i$ does not depend on the encoders of the other nodes. 

If there is a probability distribution $p(X_1, X_2, \ldots, X_n)$ on the measurements, then we can obtain a necessary and sufficient condition by considering all edge cuts.
\begin{lemma}\label{nec_suff_tree}
Consider a cut which partitions the nodes into $S$ and $\mathcal{V}\setminus S$ with $v_1 \in \mathcal{V} \setminus S$. Let $\delta^{+}(S)$ be the set of all edges from nodes in $S$ to nodes in $\mathcal{V} \setminus S$. Then the set of encoders $\{\mathcal{C}_i: 2 \leq i \leq n\}$ is feasible if and only if for every cut, the encoder on at least one of the edges in $\delta^{+}(S)$ separates $x^{1}_{S},x^{2}_{S} \in \mathcal{X}_{S}$ if there exists an assignment $x^{*}_{\mathcal{V}\setminus S}$ such that $f(x^{1}_{S}, x^{*}_{\mathcal{V}\setminus S}) \neq f(x^{2}_{S}, x^{*}_{\mathcal{V}\setminus S})$ and $p(x^{1}_{S}, x^{*}_{\mathcal{V} \setminus S})p(x^{2}_{S}, x^{*}_{\mathcal{V}\setminus S}) > 0$.
\end{lemma} 
\textbf{Proof:} Necessity is as before. For the converse, suppose the set of encoders is not feasible. Then there exist assignments $(x_1^{*}, x^{A}_{\mathcal{V} \setminus v_1})$ and $(x_1^{*}, x^{B}_{\mathcal{V} \setminus v_1})$ such that $f(x_1^{*}, x^{A}_{\mathcal{V} \setminus v_1}) \neq f(x_1^{*}, x^{B}_{\mathcal{V} \setminus v_1})$ and $p(x_1^{*}, x^{A}_{\mathcal{V} \setminus v_1})p(x_1^{*}, x^{B}_{\mathcal{V} \setminus v_1}) > 0$. However, the codewords received from nodes in $\mathcal{N}^{-}(v_1)$ are the same for both assignments. For the cut which separates $v_1$ from $\mathcal{V} \setminus v_1$, there is no encoder on $\delta^{+}(S)$ which separates $x^{A}_{\mathcal{V} \setminus v_1}$ and $x^{B}_{\mathcal{V} \setminus v_1}$. $\Box$

\begin{figure}[thpb]
\centering
\subfigure[]
{	\label{simp_net1}
	\includegraphics[width = 0.40\textwidth, height = 0.24\textheight]{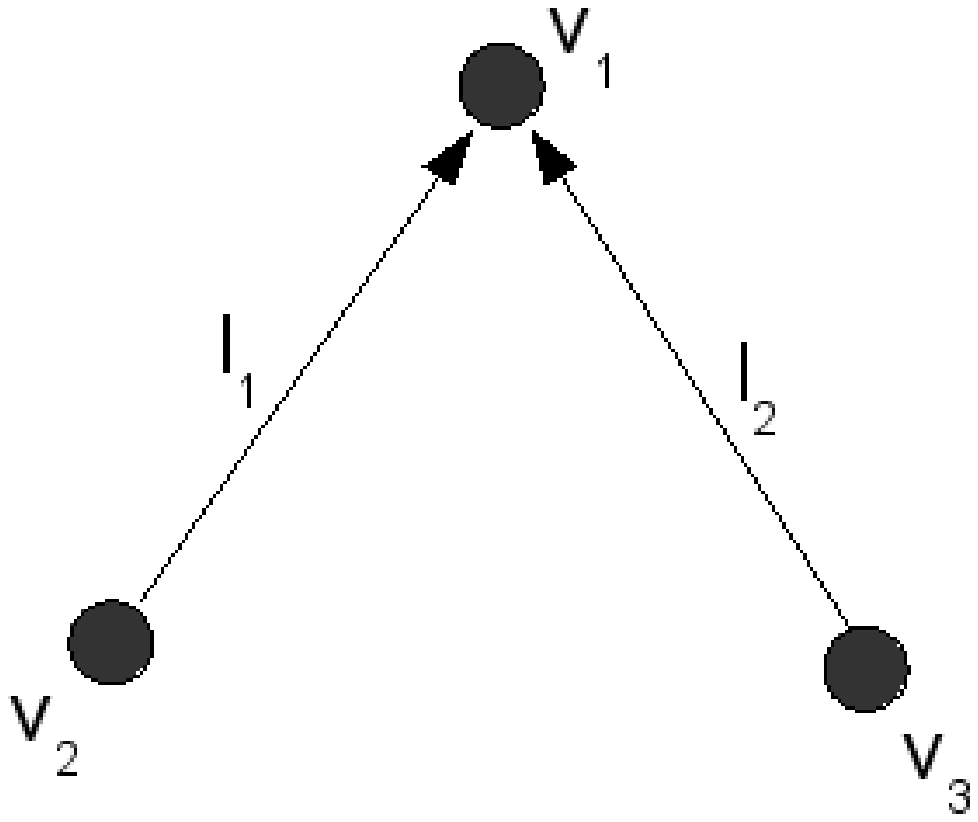}
}
\subfigure[]
{	\label{simp_net2}
	\includegraphics[width = 0.40\textwidth, height = 0.24\textheight]{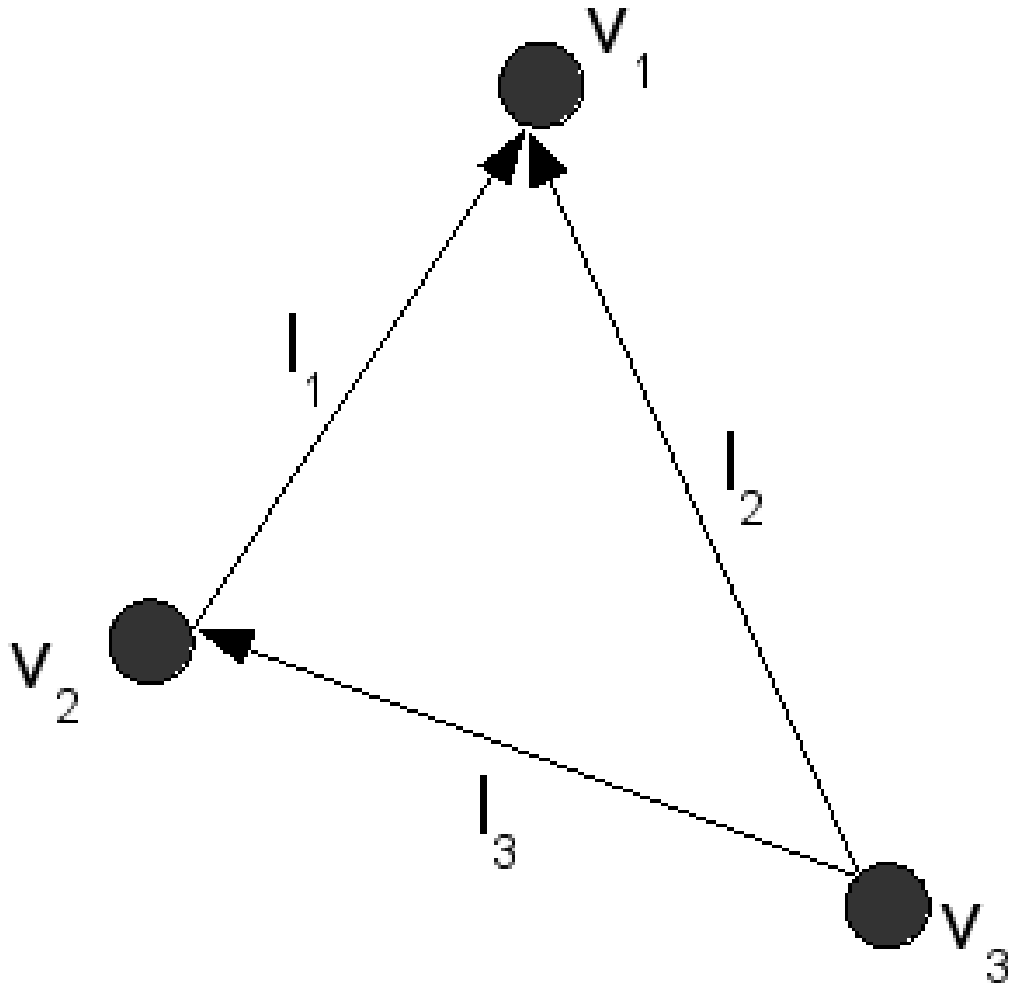}
}
\caption{Two simple networks of Examples $1$ and $2$}
\label{simp_net}
\end{figure}

The above proof of the converse is not constructive. The construction is much harder now since the encoders are coupled, as shown by the following example.
\begin{example} 
Consider the three node network $G= (\mathcal{V}, \mathcal{E})$ with $\mathcal{V} = \{v_1, v_2, v_3\}$ and $\mathcal{E} = \{(v_2, v_1), (v_3,v_1)\}$ (see Figure \ref{simp_net1}). Let $\mathcal{X}_1 = \{x^{1a}\}, \mathcal{X}_2 = \{x^{2a}, x^{2b}\}, \mathcal{X}_3 = \{x^{3a}, x^{3b}\}$. Suppose $p(x^{1a}, x^{2a}, x^{3a}) = p(x^{1a}, x^{2b}, x^{3b}) = \frac{1}{2}$. The function is given by $f(X_1, X_2, X_3) = (X_1, X_2, X_3)$. Considering the cut $(\{v_2, v_3\}, \{v_1\})$, either $v_2$ or $v_3$ needs to separate its two values. Thus the two encoders are no longer independent.
\end{example}

In general, we can trade off between the encoders on different edges. 
However, if we assume that $p(x_1, x_2, \ldots, x_n) > 0$ for all $(x_1, x_2, \ldots, x_n)$, we can separately minimize the average description length of each encoder. The optimal encoder constructs a Huffman code on the optimal alphabet $\mathcal{A}_i$. Suppose $\underline{q}_i$ is the probability vector induced on the alphabet $\mathcal{A}_i$. Then, by taking long blocks of measurements, we can achieve a minimum average case complexity arbitrarily close to $H(\underline{q}_i)$ for encoder $\mathcal{C}_i$.

\subsection{Function Computation in Directed Acyclic Graphs}\label{sec_DAG}
The extension from trees to directed acyclic graphs presents significant challenges, since there is no longer a unique path from every node to the collector. 
Consider a weakly connected directed acyclic graph (DAG) $G = (\mathcal{V}, \mathcal{E})$, where each node $v_i$ collects a block of $N$ measurements $\underline{X}_i\in \mathcal{X}_i^{N}$. The collector node $v_1$ is the unique node with only incoming edges, which wants to compute the vector function $f^{N}(\underline{X}_1, \underline{X}_2, \ldots, \underline{X}_n)$ with zero error. 

Let the encoder mapping on edge $(v_j, v_i)$ be denoted by $\mathcal{C}^{N}_{ji}$, which maps the measurement vector $\underline{X}_j$ and the codewords received thus far, to a codeword transmitted on edge $(v_j, v_i)$. Since there are no cycles in $G$, function computation proceeds in a bottom-up fashion. Node $v_i$ receives codewords $\mathcal{C}^{N}_{ji}$ on each incoming edge $(v_j, v_i)$ and then transmits a codeword $\mathcal{C}_{ik}$ on each outgoing edge $(v_i, v_k)$. A set of encoders is said to be feasible if there is a decoding function at the collector node $v_1$ which maps the received  codewords to the correct function value. Let $l_{wc}(\mathcal{C}^{N}_{ij})$ and $l_{avg}(\mathcal{C}^{N}_{ij})$ denote the worst case and average case complexity, respectively, of the encoder $\mathcal{C}^{N}_{ij}$. The rate of encoder $\mathcal{C}^{N}_{ij}$ is
\begin{displaymath}
R_{wc}(\mathcal{C}^{N}_{ij}) = \frac{l_{wc}(\mathcal{C}_{ij}^{N})}{N} \textrm{ and } R_{avg}(\mathcal{C}^{N}_{ij}) = \frac{l_{avg}(\mathcal{C}_{ij}^{N})}{N}. 
\end{displaymath}
Thus we can assign a rate vector in $\mathbf{R}^{|\mathcal{E}|}$ to every feasible set of encoders. Let $\mathcal{R}_{wc}^{(N)}$ in the worst case (or $\mathcal{R}_{avg}^{(N)}$ in the average case) be the set of feasible rate vectors for encoders of block length $N$. Then the rate region $\mathcal{R}_{wc}$ (or $\mathcal{R}_{avg}$) is given by the closure in $\mathbf{R}^{|\mathcal{E}|}$ of the finite block length rate vectors:
\begin{displaymath}
\mathcal{R}_{wc} := \overline{\bigcup_{N \geq 1} \mathcal{R}_{wc}^{(N)}} \textrm{ and } \mathcal{R}_{avg} := \overline{\bigcup_{N \geq 1} \mathcal{R}_{avg}^{(N)}}.
\end{displaymath}
Consider the following example.
\begin{example}
We have three nodes $\{v_1, v_2, v_3\}$ connected as shown in Figure \ref{simp_net2}. Let $\mathcal{X}_1 = \mathcal{X}_2 = \mathcal{X}_3 = \{0,1,2,3\}$, and suppose node $v_1$ wants to compute $f(X_1, X_2, X_3) = (X_1 + X_2 + X_3)mod 4$. It is easy to check that $(2,0,2)$ and $(2,2,0)$ are feasible rate vectors for $(l_1, l_2, l_3)$. These are rate vectors associated with the two tree subgraphs. Further, one can also check that $(2,1,1)$ is
\end{example}
\subsubsection{Outer bound on the rate region}
Consider any cut of the graph $G$ which partitions nodes into subsets $S$ and $\mathcal{V} \setminus S$ with $v_1 \in \mathcal{V} \setminus S$. Let $\delta^{+}(S)$ be the set of edges from some node in $S$ to some node in $\mathcal{V} \setminus S$. 
\begin{lemma}\label{outer_1}
Consider a set of encoders which achieve error free block function computation with rate vector $\{R_{wc}(i,j)\}_{(v_i, v_j) \in \mathcal{E}}$. Given any assignments $\underline{x}^{1}_{S}$ and $\underline{x}^{2}_{S}$ of the nodes in $S$, if there exists an assignment $\underline{x}_{\mathcal{V} \setminus S}$ such that $f^{(N)}(\underline{x}_{\mathcal{V}\setminus S}, \underline{x}^{1}_S) \neq f^{(N)}(\underline{x}_{\mathcal{V}\setminus S}, \underline{x}^{2}_S)$, then the encoders on at least one of the edges in $\delta^{+}(S)$ must separate $\underline{x}^{1}_{S}$ and $\underline{x}^{2}_{S}$. 
\begin{itemize}
\item [(i)] In the worst case block computation scenario, an outer bound on the rate region is given by
\begin{displaymath}
\sum_{(v_i, v_j) \in \delta^{+}(S)} R_{ij} \geq \log |\Pi(\mathcal{C}^{1}_{S})| \textrm{ for all cuts } (S, \mathcal{V} \setminus S),
\end{displaymath}
where $\Pi(\mathcal{C}^{1}_S)$ is the partition of $\mathcal{X}_S$ into the appropriate equivalence classes.
\item[(ii)] Suppose we have a probability distribution with $p(X_1, X_2, \ldots, X_n) > 0$. Given a cut $(S, \mathcal{V}\setminus S)$, let $R \subset \mathcal{V} \setminus S$ be the subset of nodes which have a directed path to some node in $S$. In the average case block computation scenario, an outer bound on the rate region is given by
\begin{displaymath}
\sum_{(v_i, v_j) \in \delta^{+}(S)} R_{ij} \geq H([X_S]|X_R) \textrm{ for all cuts } (S, \mathcal{V} \setminus S),
\end{displaymath}
where $[X_S]|X_R$ is the equivalence class to which $X_S$ belongs, given $X_R$ and a particular function. 
\end{itemize}
\end{lemma}
\subsubsection{Achievable region}
\begin{lemma}
Consider any directed tree subgraph $G_{T}$ with root node $v_1$. Let us suppose that only the edges in $G_T$ can be used for communication. Then we can construct encoders on each edge, which minimize worst case or average case complexity. The rate vector corresponding to a tree $G_T$ is the limit of the rate vectors for the optimal finite block length encoders for $G_T$. Thus, for a given tree $G_{T}$:
\begin{itemize}
\item[(i)] The worst case rate vector corresponding to the tree $G_T$ is an extreme point of the worst case rate region $\mathcal{R}_{wc}$ .
\item[(ii)] If $p(x_1, x_2, \ldots, x_n)>0$ for all $(x_1, x_2, \ldots, x_n)$, the rate vector corresponding to the tree $G_T$ is an extreme point of the average case rate region $\mathcal{R}_{avg}$.
\end{itemize}
\end{lemma}
The convex hull of the rate points corresponding to trees is achievable. However, we do not know if these are the \textit{only} extreme points of the rate region $\mathcal{R}$. 
\subsubsection{Some examples}
\begin{example}[Arithmetic Sum]\label{ex_arithmetic_sum}
Consider three nodes $v_1$, $v_2$, $v_3$ connected as in Figure \ref{simp_net2}. Let $\mathcal{X}_2 = \mathcal{X}_3 = \{0,1\}$, with node $v_1$ having no measurements. Suppose node $v_1$ wants to compute $f(X_1, X_2, X_3) = X_2 + X_3$. Let $(R_{21}, R_{31}, R_{32})$ be the rate vector associated with edges $(l_1, l_2, l_3)$. The outer bound on $\mathcal{R}_{wc}$ is:
\begin{displaymath}
R_{21} \geq 1; \quad R_{21} + R_{31} \geq \log 3; \quad R_{32} + R_{31} \geq 1.
\end{displaymath}
The subset of the rate region achievable by trees is: 
\begin{displaymath}
R_{21} = \lambda + (1-\lambda)\log 3, R_{31} = \lambda, R_{32} = (1-\lambda) \textrm{ for } 0 \leq \lambda \leq 1. 
\end{displaymath}
Suppose that $X_1, X_2$ are i.i.d. with $p(X_1 = 0) = p(X_1=1) = 0.5$. The outer bound on $\mathcal{R}_{avg}$ is:
\begin{displaymath}
R_{21} \geq 1; \quad R_{21} + R_{31} \geq \frac{3}{2}; \quad R_{32} + R_{31} \geq 1.
\end{displaymath}
The subset of the rate region achievable by trees is: 
\begin{displaymath}
R_{21} = \lambda + (1-\lambda)\frac{3}{2}, R_{31} = \lambda, R_{32} = (1-\lambda) \textrm{ for } 0 \leq \lambda \leq 1. 
\end{displaymath}
\end{example}
\begin{example}[Finite field parity]
Let $\mathcal{X}_i = \{0, 1, \ldots, D-1\}$ for each node $v_i$. Suppose the collector node $v_1$ wants to compute the function $(X_1 + X_2 + \ldots + X_n)\bmod{D}$. In this case, the outer bound on the worst case rate region described in Lemma \ref{outer_1} is tight.  Indeed, since the set of all outgoing links from a node is a valid cut, we have  $\sum_{(v_i, v_j) \in \mathcal{E}} R_{ij} \geq \log_{2} D$

An obvious achievable strategy is for every leaf node $v_i$ to split its block and transmit it on the outgoing edges from $v_i$. Next, we move to a node at height $1$. This node receives partial blocks from various leaf nodes, and can hence compute an intermediate parity for some instances of the block. It then splits its block along the various outgoing edges. The crucial point is that the worst case description length per instance remains $\log_{2} D$. Proceeding recursively up the DAG, we see that we can achieve the outer bound.
\end{example}
\begin{example}[Max/Min]
Let $\mathcal{X}_i = \{0, 1, \ldots, D-1\}$ for each node $v_i$. Suppose the collector node $v_1$ wants to compute $max(X_1, X_2, \ldots, X_n)$. The outer bound to the worst case rate region described in Lemma \ref{outer_1} is tight. The achievable strategy is similar to the parity case, where nodes compute intermediate maximum values and split their blocks on the outgoing edges. Once again, we utilize the fact that the range of the Max function remains constant irrespective of the number of nodes. \\
\end{example}
\section{Computing Symmetric Boolean Functions in Undirected Graphs}\label{sec_undir_graph}
In this section, we address the problem of symmetric Boolean function computation in an undirected graph. Each node has a Boolean variable  and \textit{all} nodes want to compute a given symmetric Boolean function. As in Section \ref{sec_dir_graph}, we adopt a deterministic framework and consider the problem of worst case block computation. Further, since the graph is undirected, the set of admissible strategies includes all interactive strategies, where a node may exchange several messages with other nodes, with node $i$'s transmission being allowed to depend on all previous transmissions heard by node $i$, and node $i$'s block of measurements. This is in contrast with the problem studied in Section \ref{sec_dir_graph}.

We begin by reviewing a toy problem from \cite{AhlswedeCai} where the exact communication complexity of the AND function of two variables is shown to be $\log_{2}3$ bits, for block computation. In Section \ref{sec_two_node_undir}, we generalize this approach to the two node problem, where each node $i$ has an integer variable $X_i$ and both nodes want to compute a function $f(X_1, X_2)$ which only depends on $X_1 + X_2$. We derive an optimal single-round strategy for the class of sum-threshold functions, which evaluate to 1 if $X_1 + X_2$ exceeds a threshold, and an approximate strategy for the class of sum-interval functions, which evaluate to 1 if $a \leq X_1 + X_2 \leq b$, the upper and lower bounds do not match. The general achievable strategy involves separation of the source alphabet, followed by coding, and can be used for any general function.

In Section \ref{sec_trees}, we consider symmetric Boolean function computation on trees. Since every edge is a cut-edge, we can obtain a cut-set lower bound for the number of bits that must be exchanged on an edge, by reducing it to a two node problem with general alphabets. For the class of sum-threshold functions, we are able to match the cut-set bound by constructing an achievable strategy that is reminiscent of message passing algorithms. In Section \ref{sec_gen_graphs}, for general graphs, we can still derive a cut-set lower bound by considering all partitions of the vertices. We also propose an achievable scheme that consists of activating a subtree of edges and using the optimal strategy for transmissions on the tree. While the upper and lower bounds do not match even for very simple functions, for complete graphs, we show that aggregation along trees provides a 2-OPT solution.
\subsection{The two node problem}\label{sec_two_node_undir}
Consider two nodes $1$ and $2$ with variables $X_1 \in \{0, 1, \ldots, m_1\}$ and $X_2 \in \{0, 1, \ldots, m_2\}$. Both nodes wish to compute a function $f(X_1, X_2)$ which only depends on the value of $X_1 + X_2$. To put this in context, one can suppose there are $m_1$ Boolean variables collocated at node $1$ and $m_2$ Boolean variables at node $2$, and both nodes wish to compute a symmetric Boolean function of the $n := m_1 + m_2$ variables. We pose the problem in a block computation setting, where each node $i$ has a block of $N$ independent measurements, denoted by $X_i^N$. We consider the class of all interactive strategies, where nodes $1$ and $2$ transmit messages alternately with the value of each subsequent message being allowed to depend on all previous transmissions, and the block of measurements available at the transmitting node. We define a round to include one transmission by each node. A strategy is said to achieve correct block computation if for \textit{every} choice of input $(X_1^N, X_2^N)$, each node $i$ can correctly decode the value of the function block $f^N(X_1, X_2)$ using the sequence of transmissions $b_1, b_2, \ldots$ and its own measurement block $X_i^N$. 

Let $\mathcal{S}_N$ be the set of strategies for block length $N$, which achieve zero-error block computation, and let $C(f, S_N, N)$ be the worst-case total number of bits exchanged under strategy $S_N \in \mathcal{S}_N$. The worst-case per-instance complexity of computing a function $f(X_1, X_2)$ is defined as
\begin{displaymath}
C(f) := \lim_{N \rightarrow \infty}\min_{S_N \in \mathcal{S}_N} \frac{C(f, S_N, N)}{N}.
\end{displaymath} 
\subsubsection{Complexity of sum-threshold functions}
In this paper, we are only interested in functions $f(X_1, X_2)$ which only depend on $X_1 + X_2$. Let us suppose without loss of generality that $m_1 \leq m_2$. We define an interesting class of $\{0,1\}$-valued functions called sum-threshold functions.
\begin{definition}[sum-threshold functions]
A sum-threshold function $\Pi_{\theta}(X_1, X_2)$ with threshold $\theta$ is defined as follows:
\begin{displaymath}
\Pi_{\theta}(X_1, X_2) = \left\{\begin{array}{l}1 \quad \textrm{if } X_1 + X_2 \geq \theta , \\ 0 \quad \textrm{otherwise.}\end{array} \right.
\end{displaymath}
\end{definition}
For the special case where $m_1 = 1, m_2 = 1$ and $\theta = 2$, we recover the Boolean AND function, which was studied in \cite{AhlswedeCai}. It is critical to understand this problem before we can address the general problem of computing symmetric Boolean functions. Consider two nodes with measurement blocks $X_1^N \in \{0, 1\}^N$ and $X_2^N \in \{0, 1\}^N$, which want to compute the element-wise AND of the two blocks, denoted by $\wedge^N(X_1, X_2)$.
\begin{theorem}\label{thm_two_node_and_tree}
Given any strategy $S_N$ for block computation of $X_1 \wedge X_2$, 
\begin{displaymath}
C(X_1 \wedge X_2, S_N, N) \geq N\log_{2}3.
\end{displaymath}
Further, there exists a strategy $S_N^*$ which satisfies
\begin{displaymath}
C(X_1 \wedge X_2, S_N^*, N) \leq \lceil N \log_{2}3 \rceil .
\end{displaymath}
Thus, the complexity of computing $X_1 \wedge X_2$ is given by $C(X_1 \wedge X_2)=log_{2}3$.
\end{theorem}
\textbf{Proof of achievability:} Suppose node $1$ transmits first using a prefix-free codebook. Let the length of the codeword transmitted be $l(X_1^N)$. At the end of this transmission, both nodes know the value of the function at the instances where $X_1 = 0$. Thus node $2$ only needs to indicate its bits for the instances of the block where $X_1 = 1$. Thus the total number of bits exchanged under this scheme is $l(X_1^N) + w(X_1^N)$, where $w(X_1^N)$ is the number of $1$s in $X_1^N$. For a given scheme, let us define
\begin{displaymath}
L := \max_{X_1^N}(l(X_1^N) + w(X_1^N)),
\end{displaymath} 
to be the worst case total number of bits exchanged. We are interested in finding the codebook which will result in the minimum worst-case number of bits. 

Any prefix-free code must satisfy the Kraft inequality given by $\displaystyle\sum_{X_1^N}2^{-l(X_1^N)} \leq 1$. Consider a codebook with $l(X_1^N) = \lceil N\log_{2}3\rceil - w(x_1^N)$. This satisfies the Kraft inequality since $\sum_{X_1^N}w(X_1^N) = 3^N$. Hence there exists a valid prefix free code for which the worst case number of bits exchanged is $\lceil N\log_{2}3\rceil$, which establishes that $C(X_1 \wedge X_2) \leq \log_{2}3$.

The lower bound is shown by constructing a \textit{fooling set} \cite{KushiNisan} of the appropriate size. We digress briefly to introduce the concept of fooling sets in the context of two-party communication complexity \cite{KushiNisan}. Consider two nodes $X$ and $Y$, each of which take values in finite sets $\mathcal{X}$ and $\mathcal{Y}$, and both nodes want to compute some function $f(X,Y)$ with zero error. 
\begin{definition}[Fooling Set]
A set $E \subseteq \mathcal{X} \times \mathcal{Y}$ is said to be a fooling set, if for any two distinct elements $(x_1, y_1), (x_2, y_2)$ in $E$, we have either 
\begin{itemize}
\item $f(x_1, y_1) \neq f(x_2, y_2)$, or
\item $f(x_1, y_1) = f(x_2, y_2)$, but either $f(x_1,y_2) \neq f(x_1, y_1)$ or $f(x_2,y_1) \neq f(x_1,y_1)$.
\end{itemize}
\end{definition}
Given a fooling set $E$ for a function $f(X_1, X_2)$, we have $C(f(X_1, X_2)) \geq \log_{2}|E|$. We have described two dimensional fooling sets above. The extension to multi-dimensional fooling sets is straightforward and gives a lower bound on the communication complexity of the function $f(X_1, X_2, \ldots, X_n)$. 

\textbf{Lower bound for Theorem \ref{thm_two_node_and_tree}:} We define the measurement matrix $M$ to be the matrix obtained by stacking the row $X_1^N$  over the row $X_2^N$. Thus we need to find a subset of the set of all measurement matrices which forms a fooling set. Let $E$ the set of all measurement matrices which are made up of only the column vectors {\small{$\{\left[\begin{array}{c} 1 \\ 0 \end{array}\right], \left[\begin{array}{c} 0 \\ 1 \end{array}\right], \left[\begin{array}{c} 1 \\ 1 \end{array}\right]\}$}}. We claim that $E$ is the appropriate fooling set. Consider two distinct measurement matrices $M_1, M_2 \in E$. Let $f^N(M_1)$ and $f^N(M_2)$ be the block function values obtained from these two matrices. If $f^N(M_1) \neq f^N(M_2)$, we are done. Let us suppose $f^N(M_1) = f^N(M_2)$ and since $M_1 \neq M_2$, there must exist one column where $M_1$ has $\left[\begin{array}{c} 0 \\ 1 \end{array}\right]$ but $M_2$ has $\left[\begin{array}{c} 1 \\ 0 \end{array}\right]$. Now if we replace the first row of $M_1$ with the first row of $M_2$, the resulting measurement matrix, say $M^{*}$ is such that $f(M^{*}) \neq f(M_1)$. Thus, the set $E$ is a valid fooling set. It is easy to verify that the $E$ has cardinality $3^N$. Thus, for \textit{any} strategy $S_N \in \mathcal{S}_N$, we must have $C(X_1 \wedge X_2, S_N, N) \geq N\log_{2}3$, implying that $C(X_1 \wedge X_2) \geq \log_{2}3$. This concludes the proof of Theorem \ref{thm_two_node_and_tree}. $\Box$ 

We now return to the general two node problem with $X_1 \in \{0, 1, \ldots, m_1\}$ and $X_2 \in \{0, 1, \ldots, m_2\}$ and the sum-threshold function $\Pi_{\theta}(X_1, X_2)$. We will extend the approach presented above to this general scenario.
\begin{theorem}\label{thm_sum_threshold}
Given any strategy $S_N$ for block computation of the function $\Pi_{\theta}(X_1, X_2)$,
\begin{displaymath}
C(\Pi_{\theta}(X_1, X_2), S_N, N) \geq N\log_{2}\{\min (2\theta +1, 2m_1 + 2, 2(n-\theta + 1) + 1)\}.
\end{displaymath}
Further, there exist single-round strategies $S_N^*$ and $S_N^{**}$, starting with nodes $1$ and $2$ respectively, which satisfy
\begin{displaymath}
C(\Pi_{\theta}(X_1, X_2), S_N^*, N) \leq \lceil N\log_{2}\{\min (2\theta +1, 2m_1 + 2, 2(n-\theta + 1) + 1)\} \rceil.
\end{displaymath}
\begin{displaymath}
C(\Pi_{\theta}(X_1, X_2), S_N^{**}, N) \leq \lceil N\log_{2}\{\min (2\theta +1, 2m_1 + 2, 2(n-\theta + 1) + 1)\} \rceil.
\end{displaymath}
Thus, the complexity of computing $\Pi_{\theta}(X_1, X_2)$ is given by $C(\Pi_{\theta}(X_1, X_2))=\log_{2}\{\min (2\theta +1, 2m_1 + 2, 2(n-\theta + 1) + 1)\}$.
\end{theorem}
\textbf{Proof of achievability:} We consider three cases:
\begin{itemize}
\item[(a)] Suppose $\theta \leq m_1 \leq m_2$. We specify a strategy $S_N^*$ in which node $1$ transmits first. We begin by observing that inputs $X_1 = \theta, X_1 = (\theta + 1) \ldots, X_1 = m_1$ need not be \textit{separated}, since for each of these values of $X_1$, $\Pi_{\theta}(X_1, X_2) = 1$ for all values of $X_2$. 
Thus node $1$ has an effective alphabet of $\{0, 1, \ldots, \theta\}$. Suppose node $1$ transmits using a prefix-free codeword of length $l(X_1^N)$. At the end of this transmission, node $2$ only needs to indicate one bit for the instances of the block where $X_1 = 0, 1, \ldots, (\theta -1)$. Thus the worst-case total number of bits is 
\begin{displaymath}
L := \max_{X_1^N} (l(X_1^N) + w^{0}(X_1^N) + w^{1}(X_1^N) + \ldots + w^{\theta -1}(X_1^N)),
\end{displaymath}
where $w^{j}(X_1^N)$ is the number of instances in the block where $X_1 = j$. We are interested in finding the codebook which will result in the minimum worst-case number of bits. 
From the Kraft inequality for prefix-free codes we have
\begin{equation}
\sum_{X_1^N \in \{0, 1, \ldots, \theta\}^N}2^{-L + w^{0}(X_1^N) + w^{1}(X_1^N) + \ldots + w^{\theta -1}(X_1^N))} \leq 1. \nonumber
\end{equation}  
Consider a codebook with $l(X_1^N) = \lceil N\log_{2}(2\theta +1)\rceil - w(x_1^N)$. This satisfies the Kraft inequality since 
\begin{displaymath}
\sum_{X_1^N \in \{0, 1, \ldots, \theta\}^N}2^{w^{0}(X_1^N) + w^{1}(X_1^N) + \ldots + w^{\theta -1}(X_1^N))}.1^{w^{\theta}(X_1^N)} = (2\theta + 1)^N.
\end{displaymath}
Hence there exists a prefix-free code for which the worst-case total number of bits exchanged is $\lceil N\log_{2}(2\theta +1) \rceil$. Since $\theta \leq m_1 \leq m_2$, we have
\begin{displaymath}
C(\Pi_{\theta}(X_1, X_2), S_N^*, N) \leq \lceil N\log_{2}\{\min (2\theta +1, 2m_1 + 2, 2(n-\theta + 1) + 1)\}\rceil.
\end{displaymath}
The strategy $S_N^{**}$ starting at node $2$ can be similarly derived. Node $2$ now has an effective alphabet of $\{0, 1, \ldots, \theta\}$, and we have $C(\Pi_{\theta}(X_1, X_2), S_N^{**}, N) \leq \lceil N\log_{2} (2\theta +1) \rceil$.\\
\item[(b)] Suppose $m_1 \leq m_2 < \theta$. Consider a strategy $S_N^*$ in which node $1$ transmits first. The inputs $X_1 = 0, X_1 = 1, \ldots, X_1 = \theta - m_2 -1$ need not be \textit{separated} since for each of these values of $X_1$, $\Pi_{\theta}(X_1, X_2) = 0$ for all values of $X_2$. Thus node $1$ has an effective alphabet of $\{\theta - m_2 - 1, \theta - m_2, \ldots, m_1\}$. Upon hearing node $1$'s transmission, node $2$ only needs to indicate one bit for the instances of the block where $X_1 = \theta - m_2, \ldots, m_1$. Consider a codebook with $l(X_1^N) = \lceil N\log_{2}(2(m_1 + m_2 - \theta +1) + 1)\rceil - w^{\theta - m_2}(X_1^N) - \ldots - w^{m_1}(X_1^N)$. This satisfies the Kraft inequality and we have $L = \lceil N \log_{2}(2(n- \theta + 1) + 1)\rceil$. Since $m_1 \leq m_2 < \theta$, we have that
\begin{displaymath}
C(\Pi_{\theta}(X_1, X_2), S_N^*, N) \leq \lceil N\log_{2}\{\min (2\theta +1, 2m_1 + 2, 2(n-\theta + 1) + 1)\}\rceil .
\end{displaymath}
The strategy $S_N^{**}$ starting at node $2$ can be analogously derived. \\
\item[(c)] Suppose $m_1 < \theta \leq m_2$. For the case where node $1$ transmits first, we construct a trivial strategy $S_N^*$ where node $1$ uses a codeword of length $\lceil N\log_{2}(m_1 + 1)\rceil$ bits and node $2$ replies with a string of $N$ bits indicating the function block. Thus we have $C(\Pi_{\theta}(X_1, X_2), S_N^*, N) \leq \lceil N\log_{2}(2m_1 + 2)\rceil$. 
\end{itemize}

Now consider a strategy $S_N^{**}$ where node $2$ transmits first. Observe that the inputs $X_2 = 0, X_2 = 1, \ldots, X_2 = \theta - m_1 -1$ need not be separated since for each of these values of $X_2$, $\Pi_{\theta}(X_1, X_2) = 0$ for all values of $X_2$. Further, the inputs $X_2 = \theta , X_2 = \theta + 1, \ldots, X_2 = m_2$ need not be \textit{separated}. Thus node $1$ has an effective alphabet of $\{\theta - m_1 - 1, \theta - m_1, \ldots, \theta\}$. Upon hearing node $2$'s transmission, node $1$ only needs to indicate one bit for the instances of the block where $X_2 = \theta - m_1, \ldots, \theta - 1$. Consider a codebook with $l(X_2^N) = \lceil N\log_{2}(2m_1 + 2)\rceil - w^{\theta - m_1}(X_1^N) - \ldots - w^{\theta -1}(X_1^N)$. This satisfies the Kraft inequality and we have $L = \lceil N\log_{2}(2(n - \theta +1) + 1) \rceil$. Since $m_1 < \theta \leq m_2$, we have that
\begin{displaymath}
C(\Pi_{\theta}(X_1, X_2), S_N^{**}, N) \leq \lceil N\log_{2}\{\min (2\theta +1, 2m_1 + 2, 2(n-\theta + 1) + 1)\}\rceil .
\end{displaymath}
The lower bound is shown by constructing a \textit{fooling set} as before. Let $E$ denote the set of all measurement matrices which are made up only of the column vectors from the set 
\begin{displaymath}
Z = \left\{\left[\begin{array}{c} z_1 \\ z_2 \end{array}\right]: 0 \leq z_1 \leq m_1, 0 \leq z_2 \leq m_2, (\theta -1) \leq z_1 + z_2 \leq \theta \right\}.
\end{displaymath}

We claim that $E$ is the appropriate fooling set. Consider two distinct measurement matrices $M_1, M_2 \in E$. Let $f^N(M_1)$ and $f^N(M_2)$ be the block function values obtained from these two matrices. If $f^N(M_1) \neq f^N(M_2)$, we are done. Let us suppose $f^N(M_1) = f^N(M_2)$, and note that since $M_1 \neq M_2$, there must exist one column where $M_1$ and $M_2$ differ. Suppose $M_1$ has $\left[\begin{array}{c} z_{1a} \\ z_{2a} \end{array}\right]$ while $M_2$ has $\left[\begin{array}{c} z_{1b} \\ z_{2b} \end{array}\right]$, where $z_{1a} + z_{2a} = z_{1b} + z_{2b}$. Assume without loss of generality that $z_{1a} < z_{1b}$ and $z_{2a} > z_{2b}$. 
\begin{itemize}
\item If $z_{1a} + z_{2a} = z_{1b} + z_{2b} = \theta -1$, then the \textit{diagonal} element $f(z_{1b}, z_{2a}) = 1$ since $z_{1b} + z_{2a} \geq \theta$. Thus, if we replace the first row of $M_1$ with the first row of $M_2$, the resulting measurement matrix, say $M^{*}$, is such that $f(M^{*}) \neq f(M_1)$.
\item If $z_{1a} + z_{2a} = z_{1b} + z_{2b} = \theta$, then the \textit{diagonal} element $f(z_{1a}, z_{2b}) = 0$ since $z_{1b} + z_{2a} < \theta$. Thus, if we replace the second row of $M_1$ with the second row of $M_2$, the resulting matrix $M^{*}$ is such that $f(M^{*}) \neq f(M_1)$.
\end{itemize}
Thus, the set $E$ is a valid fooling set with cardinality $|Z|^N$. For any strategy $S_N$, we have $C(f, S_N, N) \geq N\log_{2}|Z|$. The cardinality of $Z$ can be modeled as the sum of the coefficients of $Y^{\theta}$ and $Y^{\theta -1 }$ in a carefully constructed polynomial:
\begin{eqnarray}
|Z| & = & \left[Y^{\theta}\right] + \left[Y^{\theta -1 }\right](1 + Y + \ldots + Y^{m_1})(1 + Y + \ldots + Y^{m_2}) \nonumber \\
& = & \left[Y^{\theta}\right] + \left[Y^{\theta -1 }\right] \frac{(1 - Y^{m_1 + 1})(1 - Y^{m_2 + 1})}{(1-Y)^2}. \nonumber
\end{eqnarray}
This is solved using the binomial expansion for $\frac{1}{(1-Y)^k}$ \cite{West}.
\begin{displaymath}
|Z| = \left[Y^{\theta}\right] + \left[Y^{\theta -1 }\right] (1 - Y^{m_1 + 1})(1 - Y^{m_2 + 1})\sum_{k = 0}^{\infty}\left(\begin{array}{c}k+1 \\ 1\end{array}\right)Y^k .
\end{displaymath}
\begin{itemize}
\item[(a)] Suppose $\theta \leq m_1 \leq m_2$. Then $|Z| = \theta + \theta + 1$.
\item[(b)] Suppose $m_1 \leq \theta \leq m_2$. Then $|Z| = 2m_1 + 2$.
\item[(c)] Suppose $m_1 \leq m_2 \leq \theta$. Then $|Z| = 2(n- \theta + 1) +1$.
\end{itemize}
This completes the proof of Theorem \ref{thm_sum_threshold}. $\Box$ 
\subsubsection{Complexity of sum-interval functions}
\begin{definition}[sum-interval functions]
A sum-interval function $\Pi_{[a,b]}(X_1, X_2)$ on the interval $[a,b]$ is defined as follows:
\begin{displaymath}
\Pi_{[a,b]}(X_1, X_2) := \left\{\begin{array}{l}1 \quad \textrm{if } a \leq X_1 + X_2 \leq b, \\ 0 \quad \textrm{otherwise.} \end{array}\right.
\end{displaymath}
\end{definition}
\begin{theorem}\label{thm_sum_interval}
Given any strategy $S_N$ for block computation of $\Pi_{[a,b]}(X_1, X_2)$ where $b \leq n/2$,
\begin{displaymath}
C(\Pi_{[a,b]}(X_1, X_2), S_N, N) \geq N\log_{2}\{\min (2b-a+3, m_1 + 1)\}.
\end{displaymath}
Further, there exists a single-round strategy $S_N^*$ which satisfies
\begin{displaymath}
C(\Pi_{[a,b]}(X_1, X_2), S_N^*, N) \leq \lceil N\log_{2}\{\min (2(b+1)+1, 2m_1 + 2)\} \rceil .
\end{displaymath}
Thus, we have obtained the complexity of computing $\Pi_{\theta}(X_1, X_2)$ to within one bit.
\end{theorem}
\textbf{Proof of Achievability:}
\begin{itemize}
\item[(a)] Suppose $b \leq m_1 \leq m_2$. Node $1$ has an effective alphabet of $\{0, 1, \ldots, b+1\}$. Then the worst-case total number of bits exchanged is given by
\begin{displaymath}
L := \max_{X_1^N}(l(X_1^N) + w^{0}(X_1^N) + \ldots + w^{b}(X_1^N)).
\end{displaymath}
From the Kraft inequality, we can obtain a prefix free codebook with $L = \lceil N\log_{2}(2b + 1) + 1)\rceil$. Thus we have 
\begin{displaymath}
C(\Pi_{[a,b]}(X_1, X_2), S_N^*, N) \leq \lceil N\log_{2}(2(b+1) + 1) \rceil.
\end{displaymath}
\item[(b)] Suppose $m_1 \leq a \leq b \leq m_2$ or $a \leq m_1 \leq b \leq m_2$. In either of these scenarios, node $1$ has an effective alphabet of $\{0, 1, \ldots, m_1\}$. Then the worst-case total number of bits exchanged is given by 
\begin{displaymath}
L := \max_{X_1^N}(l(X_1^N) + w^{a-m_2}(X_1^N) + \ldots + w^{m_1}(X_1^N))
\end{displaymath}
From the Kraft inequality, we can obtain a prefix free codebook with $L = \lceil N\log_{2}(2m_1 + 2)\rceil$. Thus we have $C(\Pi_{[a,b]}(X_1, X_2), S_N^*, N) \leq \lceil N\log_{2}(2m_1 + 2) \rceil$.
\end{itemize}
\textbf{Proof of Lower Bound:} We attempt to find a fooling subset $E$ of the set of measurement matrices. Our first guess would be the set of measurement matrices which are composed of only column vectors which sum up to $b$ or $b+1$. However we see that this is not necessarily a fooling set, because if $[z_{1a}, z_{2a}]^T$ and $[z_{1b}, z_{2b}]^T$ are two columns which sum to $b+1$, and if $z_{1a} \leq z_{1b} -(b-a+2)$, then neither of the diagonal elements evaluate to function value $1$. Thus, we can pick a maximum of $(b-a+2)$ consecutive elements along the line $z_1 + z_2 = b+1$, and, as before, all the elements on the line $z_1 + z_2 = b$. It is easy to check that this modified set of columns indeed yields a fooling set of measurement matrices. Now we need to compute the number of such columns.
\begin{itemize}
\item[(a)] Suppose $b \leq m_1 \leq m_2$. The number of columns which sum up to $b$ is equal to $b+1$. Thus the size of the fooling set is given by $(2b -a +3)^N$. 
\item[(b)] Suppose $a \leq m_1 \leq b \leq m_2$ or $m_1 \leq a \leq b \leq m_2$. The number of columns which sum up to $b$ is equal to $m_1 + 1$ and the number of columns which sum up to $b+1$ is equal to $m_1 + 1$. Thus, the size of the fooling set is given by $\{(m_1 + 1) + \min(m_1 + 1, b-a +2)\}^N$.
\end{itemize}
\subsubsection{A general strategy for achievability}
The strategy for achievability used in Theorems \ref{thm_sum_threshold} and \ref{thm_sum_interval} suggests an achievable scheme for any general function $f(X_1, X_2)$ of variables $X_1 \in \mathcal{X}_1$ and $X_2 \in \mathcal{X}_2$ which depends only on the value of $X_1 + X_2$. This is done in two stages.
\begin{itemize}
\item \textbf{Separation:} Two inputs $x_{1a}$ and $x_{1b}$ need not be \textit{separated} if $f(x_{1a}, x_2) = f(x_{1b}, x_2)$ for all values $x_2$. By checking this condition for each pair $(x_{1a}, x_{1b})$, we can arrive at a partition of $\{0, 1 \ldots, m_1\}$ into equivalence classes, which can be considered a reduced alphabet, say $A := \{a_1, \ldots, a_l\}$. 
\item \textbf{Coding:} Let $A_{0}$ denote the subset of the alphabet $A$ for which the function evaluates only to $0$, irrespective of the value of $X_2$, and let $A_{1}$ denote the subset of $A$ which always evaluates to $1$. Clearly, from the equivalence class structure, we have $|A_{0}| \leq 1$ and $|A_{1}| \leq 1$. Using the Kraft inequality as in Theorems \ref{thm_sum_threshold} and \ref{thm_sum_interval}, we obtain a scheme $S_N^*$ with complexity $\log_{2}(2l - |A_0| - |A_1|)$.
\end{itemize}
\subsection{Computing symmetric Boolean functions on tree networks}\label{sec_trees}
Consider a tree graph $T = (V, E)$, with node set $V = \{0, 1, \ldots, n\}$ and edge set $E$. Each node $i$ has a Boolean variable $X_i \in \{0,1\}$, and every node wants to compute a given symmetric Boolean function $f(X_1, X_2, \ldots, X_n)$. Again, we allow for block computation and consider all strategies where nodes can transmit in any sequence with possible repetitions, subject to:
\begin{itemize}
\item On any edge $e = (i,j)$, either node $i$ transmits or node $j$ transmits, or neither, and this is determined from the previous transmissions.
\item Node $i$'s transmission can depend on the previous transmissions and the measurement block $X_i^N$.
\end{itemize} 
For sum-threshold functions, we have a computation and communication strategy that is optimal for each link. 
\begin{theorem}\label{thm_tree_threshold}
Consider a tree network where we want to compute the function $\Pi_{\theta}(X_1, \ldots, X_n)$. Let us focus on a single edge $e \equiv (i,j)$ whose removal disconnects the graph into components $A_{e}$ and $V\setminus A_{e}$, with $|A_{e}| \leq |V \setminus A_{e}|$. For any strategy $S_N \in \mathcal{S}_N$, the number of bits exchanged along edge $e \equiv (i,j)$, denoted by $C_{e}(\Pi_{\theta}(X_1, \ldots, X_n), S_N, N)$, is lower bounded by
\begin{equation}
C_{e}(\Pi_{\theta}(X_1, \ldots, X_n), S_N, N) \geq N\log_{2}\{\min (2\theta +1, 2|A_{e}| + 2, 2(n-\theta + 1) + 1)\}. \nonumber
\end{equation}
Further, there exists a strategy $S_N^*$ such that for any edge $e$,
\begin{equation}
C_{e}(\Pi_{\theta}(X_1, \ldots, X_n), S_N^*, N) \leq \lceil N\log_{2}\{\min (2\theta +1, 2|A_{e}| + 2, 2(n-\theta + 1) + 1)\}\rceil. \nonumber
\end{equation}
The complexity of computing $\Pi_{\theta}(X_1, \ldots, X_n)$ is given by
\begin{displaymath}
C_{e}(\Pi_{\theta}(X_1, \ldots, X_n)) = \log_{2}\{\min (2\theta +1, 2|A_{e}| + 2, 2(n-\theta + 1) + 1)\}.
\end{displaymath}
\end{theorem}
\textbf{Proof:} Given a tree network $T$, every edge $e$ is a cut edge. Consider an edge $e$ whose removal creates components $A_{e}$ and $V \setminus A_{e}$, with $|A_{e}| \leq  |V \setminus A_{e}|$. Now let us aggregate the nodes in $A_{e}$ and also those in $V \setminus A_{e}$, and view this as a problem with two nodes connected by edge $e$. Clearly the complexity of computing the function $\Pi_{\theta}(X_{A_{e}}, X_{V \setminus A_{e}})$ is a lower bound on the worst-case total number of bits that must be exchanged on edge $e$ under any strategy $S_N$. Hence we obtain
\begin{equation}
C_{e}(\Pi_{\theta}(X_1, \ldots, X_n), S_N, N) \geq N\log_{2}\{\min (2\theta +1, 2|A_{e}| + 2, 2(n-\theta + 1) + 1)\}. \nonumber
\end{equation}
The achievable strategy $S_N^*$ is derived from the achievable strategy for the two node case in Theorem \ref{thm_sum_threshold}. While the transmissions back and forth along any edge will be exactly the same, we need to orchestrate these transmissions so that conditions of causality are maintained. Pick any node, say $r$, to be the root. This induces a partial order on the tree network. We start with each leaf in the network transmitting its codeword to the parent. Once a parent node obtains a codeword from each of its children, it has sufficient knowledge to disambiguate the letters of the effective alphabet of the subtree, and subsequently it transmits a codeword to its parent. Thus codewords are transmitted from child nodes to parent nodes until the root is reached. The root can then compute the value of the function and now sends the appropriate replies to its children. The children then compute the function and send appropriate replies, and so on. This sequential strategy  depends critically on the fact that, in the two node problem, we derived optimal strategies starting from either node. For any edge $e$, the worst-case total number of bits exchanged is given by
\begin{equation}
C_{e}(\Pi_{\theta}(X_1, \ldots, X_n), S_N^*, N) \leq \lceil N\log_{2}\{\min (2\theta +1, 2|A_{e}| + 2, 2(n-\theta + 1) + 1)\}\rceil . \Box \nonumber
\end{equation}
One can similarly derive an approximately optimal strategy for sum-interval functions, which we state here without proof.
\begin{theorem} \label{thm_tree_interval}
Consider a tree network where we want to compute the function $\Pi_{[a,b]}(X_1, \ldots, X_n)$, with $b \leq \frac{n}{2}$. Let us focus on a single edge $e \equiv (i,j)$ whose removal disconnects the graph into components $A_{e}$ and $V\setminus A_{e}$, with $|A_{e}| \leq |V \setminus A_{e}|$. For any strategy $S_N \in \mathcal{S}_N$, the number of bits exchanged along edge $e \equiv (i,j)$, denoted by $C_{e}(f, S_N, N)$ is lower bounded by 
\begin{displaymath}
C_{e}(\Pi_{[a,b]}(X_1, \ldots, X_n), S_N, N) \geq N\log_{2}\{\min (2b-a+3, |A_e| + 1)\}. 
\end{displaymath}
Further there exists a strategy $S_N^*$ such that for any edge $e$,
\begin{displaymath}
C_{e}(\Pi_{[a,b]}(X_1, \ldots, X_n), S_N^*, N) \leq \lceil N\log_{2}\{\min (2(b+1) +1, 2|A_e| + 2)\}\rceil.
\end{displaymath}
\end{theorem}
\subsection{Extension to non-binary alphabets}
The extension to the case where each node draws measurements from a non-binary alphabet is immediate. 
Consider a tree network with $n$ nodes where node $i$ has a measurement $X_i \in \{0, 1, \ldots, l_i -1\}$. Suppose all nodes want to compute a given function which only depends on the value of $X_1 + X_2 + \ldots + X_n$. We can define sum-threshold functions in analogous fashion and derive an optimal strategy for computation. 
\begin{theorem}
Consider a tree network where we want to compute a sum-threshold function, $\Pi_{\theta}(X_1, \ldots, X_n)$, of non-binary measurements. Let us focus on a single edge $e$ whose removal disconnects the graph into components $A_{e}$ and $V\setminus A_{e}$. Let us define $l_{A_e} := \sum_{i \in A_e} l_i$. Then the complexity of computing $\Pi_{\theta}(X_1, \ldots, X_n)$ is given by 
\begin{equation}
C_{e}(\Pi_{\theta}(X_1, \ldots, X_n)) \\
= \log_{2}\{\min (2\theta +1, 2\min(l_{A_e}, l_{V \setminus A_e}) + 2, 2(l_{V}-\theta + 1) + 1)\}. \nonumber
\end{equation}
\end{theorem}
Theorem \ref{thm_tree_interval} also extends to the case of non-binary alphabets.
\subsection{Computing sum-threshold functions in general graphs}\label{sec_gen_graphs}
We now consider the computation of sum-threshold functions in general graphs where the alphabet is not restricted to be binary. A cut is defined to be a set of edges $F \subseteq E$ which disconnect the network into two components $A_F$ and $V \setminus A_F$. 
\begin{lemma}[Cut-set bound]
Consider a general network $G = (V, E)$, where node $i$ has measurement $X_i \in \{0, 1, \ldots, l_i - 1\}$ and all nodes want to compute the function $\Pi_{\theta}(X_1, \ldots, X_n)$. Given a cut $F$ which separates $A_F$ from $V \setminus A_F$, the cut-set lower bound specifies that: For any strategy $S_N$, the number of bits exchanged on the edges in $F$ is lower bounded by
\begin{equation}
C_{F}(\Pi_{\theta}(X_1, \ldots, X_n), S_N, N) \geq N \log_{2}(\min \{2\theta +1, 2m_F + 2, 2(l_{V} - \theta + 1) + 1)\}. \nonumber
\end{equation}
where $l_{A_F} = \sum_{i \in A_F} l_i$ and $m_F = \min(l_{A_F}, l_{V \setminus A_F})$.
\end{lemma}

A natural achievable strategy is to pick a spanning subtree of edges and use the optimal strategy on this subtree. The convex hull of the rate vectors of the subtree aggregation schemes, is an achievable region. We wish to compare this with the cut-set region. To simplify matters, consider a complete graph $G$ where each node $i$ has a measurement $X_i \in \{0, \ldots, l-1\}$. Let $R_{ach}$ be the maximum symmetric ratepoint achievable by aggregating along trees, and $R_{cut}$ be the minimum symmetric ratepoint that satisfies the cut-set constraints. 
\begin{theorem} 
For the computation of sum-threshold functions on complete graphs, $R_{ach} \leq 2(1 - \frac{1}{n}))R_{cut}$. In fact, this approximation ratio is tight.
\end{theorem}
\textbf{Proof:} Let us assume without loss of generality that $\theta \leq \frac{n.l}{2}$. Consider all cuts of the type $(\{i\}, V \setminus \{i\})$. This yields 
\begin{displaymath}
R_{cut} \geq \max_{i \in V}\left(\frac{\min(\log_{2}(2\theta + 1), \log_{2}(2l_i+2))}{n-1}\right).
\end{displaymath}
Now consider the achievable scheme which employs each of the $n$ star graphs for equal sized sub-blocks of measurements. 
The rate on edge $(i,j)$ is given by
\begin{displaymath}
\frac{1}{n}\left(\min(\log_{2}(2\theta + 1), \log_{2}(2l_i+2)) + \min(\log_{2}(2\theta + 1), \log_{2}(2l_j+2))\right)
\end{displaymath}
Hence we have
\begin{equation}
R_{ach} \leq \frac{2}{n}(\min(\log_{2}(2\theta + 1), \max_{i \in V}\{\log_{2}(2l_i+2)\})) \leq 2\left(1 - \frac{1}{n}\right)R_{cut}. \nonumber
\end{equation}
\textbf{Tight Example:} Suppose $l_1 = l_2 = \ldots = l_n = l$ and $\theta > l$, then 
\begin{displaymath}
R_{cut} = \frac{1}{n-1}\min(\log_{2}(2\theta + 1), \log_{2}(2l+2))
\end{displaymath}
Further, from the symmetry of the problem, it is clear that the optimal scheme is to employ the $n$ star graphs for equal sub-blocks of measurements. This gives a symmetric achievable point of
\begin{displaymath}
R_{ach} = \frac{2}{n}\min(\log_{2}(2\theta + 1), \log_{2}(2l+2)) = 2\left(1 - \frac{1}{n}\right)R_{cut}.
\end{displaymath}
\subsection{Linear Programming Formulation}
The above approach of restricting attention to aggregation along star graphs, gives in to a convenient Linear Programming (LP) formulation. Consider a complete graph $G$. Let us define the rate region achievable by star graphs in the following way
\begin{displaymath}
\tilde{\mathcal{R}}_{ach} = \{ A\underline{\lambda}: ||\lambda ||_{1} = 1\}
\end{displaymath}
where $A$ is a $n \times \frac{n(n-1)}{2}$ matrix where $a_{ie}$th entry is the minimum number of bits that must be sent along edge $e$ under tree aggregation scheme $T_i$. The vector $\underline{\lambda}$ is the relative weights assigned to the different trees. We want to compare the rate vectors achieved by this scheme with the rate vectors that satisfy the cut constraints. Let $\underline{r} \in \mathcal{R}_{cut}$ be a given rate vector which satisfies the cut constraints of Lemma 1. Now, we seek to find an achievable rate vector that is within a $\theta$ factor of $\underline{r}$, and further, we want to find the minimum value of such a $\theta$. This can formulated as a linear program
\begin{center}
Min. $\theta$  \\
s.t. $A\underline{\lambda} \leq \theta \underline{r}$ \\
$\quad ||\lambda ||_{1} \geq 1$ \\
$\quad \lambda \geq 0$, $\theta \geq 0$
\end{center}
Thus we can obtain the optimal assignment $\lambda ^{*}$ and the optimal factor $\theta ^ {*}$. Note that this assignment depends on the given rate vector $\underline{r} \in \mathcal{R}_{cut}$. We can also write similar such LPs for other classes of trees.
\section{Concluding remarks}
In this paper, we have addressed some problems that arise in the context of information aggregation in sensor networks. While the general problem of devising optimal strategies for function computation in wireless networks appears formidable, we have simplified it by abstracting out the medium access control problem and analyzing the problem of function computation in graphs. 

We have started with the problem of zero error function computation in directed graphs, and analyzed both worst case and average case metrics. For directed tree graphs, we have constructed optimal encoding schemes on each edge. This matches the cut-set lower bounds.  For general DAGs, we have provided an outer bound on the rate region, and an achievable region based on aggregating along subtrees. While we have presented some examples where tree aggregation schemes are optimal, it remains to quantify the sub-optimality of tree aggregation schemes in general. 

We have also addressed the computation of symmetric Boolean functions in undirected graphs, where all nodes want to compute the function. For the case of computing sum-threshold functions in undirected trees, we have derived the optimal strategy for each edge. The achievable scheme for block computation involves a layering of transmissions that is reminiscent of message passing. Our framework can be generalized to handle functions of integer measurements which only depend on the sum of the measurements. The extension to general graphs is very interesting and appears significantly harder. However, a cut-set lower bound can be immediately derived, and in some special cases one can show that subtree aggregation schemes provide a 2-OPT solution. Once again, it remains to study the suboptimality of tree aggregation schemes in general graphs.
\singlespacing
\bibliographystyle{unsrt}
\bibliography{defense_biblio}

\begin{thebibliography}{10}

\bibitem{AhlswedeYeung}
R.~Ahlswede, N.~Cai, S.~R. Li, and R.~W. Yeung.
\newblock Network information flow.
\newblock {\em IEEE Transactions on Information Theory}, 46(4):1204--1216, July
  2000.

\bibitem{AppuFran}
R.~Appuswamy, M.~Franceschetti, N.~Karamchandani, and K.~Zeger.
\newblock Network coding for computing.
\newblock In {\em Proceedings of the 46th Annual Allerton Conference on
  Communication, Control, and Computing}, pages 1--6, September 2008.

\bibitem{GiridharKumar}
A.~Giridhar and P.~R. Kumar.
\newblock Computing and communicating functions over sensor networks.
\newblock {\em IEEE Journal on Selected Areas in Communication},
  23(4):755--764, April 2005.

\bibitem{GuptaShak}
S.~Subramanian, P.~Gupta, and S.~Shakkottai.
\newblock Scaling bounds for function computation over large networks.
\newblock In {\em Proceedings of the IEEE International Symposium on
  Information Theory (ISIT)}, pages 136--140, June 2007.

\bibitem{KushiNisan}
E.~Kushilevitz and N.~Nisan.
\newblock {\em Communication Complexity}.
\newblock Cambridge University Press, New York, NY, USA, 1997.

\bibitem{Wegener}
I.~Wegener.
\newblock {\em The {C}omplexity of {B}oolean {F}unctions}.
\newblock J. Wiley \& Sons, Inc., New York, NY, USA, 1987.

\bibitem{OrlitskyElgamal}
A.~Orlitsky and A.~El Gamal.
\newblock Average and randomized communication complexity.
\newblock {\em IEEE Transactions on Information Theory}, 36:3--16, 1990.

\bibitem{KarchmerRazWigderson}
M.~Karchmer, R.~Raz, and A.~Wigderson.
\newblock Super-logarithmic depth lower bounds via direct sum in communication
  coplexity.
\newblock In {\em Structure in Complexity Theory Conference}, pages 299--304,
  1991.

\bibitem{AhlswedeCai}
R.~Ahlswede and Ning Cai.
\newblock On communication complexity of vector-valued functions.
\newblock {\em IEEE Transactions on Information Theory}, 40:2062--2067, 1994.

\bibitem{Kschischang}
F.~R. Kschischang, B.~J. Frey, and H.~Loeliger.
\newblock Factor graphs and the sum-product algorithm.
\newblock {\em IEEE Transactions on Information Theory}, 47(2):498--519,
  February 2001.

\bibitem{AjiMceliece}
S.~Aji and R.~Mceliece.
\newblock The generalized distributive law.
\newblock {\em IEEE Transactions on Information Theory}, 46(2):325--343, 2000.

\bibitem{WynerZiv}
A.~D. Wyner and J.~Ziv.
\newblock The rate-distortion function for source coding with side information
  at the decoder.
\newblock {\em IEEE Transactions on Information Theory}, 22(1):1--10, January
  1976.

\bibitem{OrlitskyRoche}
A.~Orlitsky and J.~R. Roche.
\newblock Coding for computing.
\newblock {\em IEEE Transactions on Information Theory}, 47:903--917, 2001.

\bibitem{Witsenhausen}
H.~Witsenhausen.
\newblock The zero-error side information problem and chromatic numbers.
\newblock {\em IEEE Transactions on Information Theory}, 22:592--593, September
  1976.

\bibitem{AlonOrlitsky}
N.~Alon and A.~Orlitsky.
\newblock Source coding and graph entropies.
\newblock {\em IEEE Transactions on Information Theory}, 42:1329--1339,
  September 1996.

\bibitem{MaIshwar}
N.~Ma and P.~Ishwar.
\newblock Two-terminal distributed source coding with alternating messages for
  function computation.
\newblock In {\em Proceedings of the IEEE International Symposium on
  Information Theory (ISIT)}, pages 51--55, 2008.

\bibitem{MaGuptaIshwar}
N.~Ma, P.~Ishwar, and P.~Gupta.
\newblock Information-theoretic bounds for multiround function computation in
  collocated networks.
\newblock In {\em Proceedings of the IEEE International Symposium on
  Information Theory (ISIT)}, pages 2306--2310, 2009.

\bibitem{Gallager}
R.~D. Gallager.
\newblock Finding parity in a simple broadcast network.
\newblock {\em IEEE Transactions on Information Theory}, 34(2):176--180, March
  2008.

\bibitem{YingSrikantDullerud}
L.~Ying, R.~Srikant, and G.~E. Dullerud.
\newblock Distributed symmetric function computation in noisy wireless sensor
  networks.
\newblock {\em IEEE Transactions on Information Theory}, 53(12):4826--4833,
  December 2007.

\bibitem{Dutta}
C.~Dutta, Y.~Kanoria, D.~Manjunath, and J.~Radhakrishnan.
\newblock A tight lower bound for parity in noisy communication networks.
\newblock In {\em Proceedings of the 20th ACM-SIAM Symposium on Discrete
  Algorithms (SODA)}, pages 1056--1065, January 2008.

\bibitem{West}
D.~West.
\newblock {\em Combinatorial Mathematics}.
\newblock Course notes for ECE 580, Department of Mathematics, University of
  Illinois at Urbana-Champaign, 2008.

\end{thebibliography}
\end{document}